\documentclass[10pt,journal,twoside]{IEEEtran}
\usepackage{amsmath}
\usepackage{graphicx}
\usepackage{amsfonts}
\usepackage{amssymb}
\usepackage{psfrag}
\usepackage{cite}
\usepackage{color}
\usepackage{multirow}
\usepackage{subfigure}
\usepackage[linesnumbered,lined,commentsnumbered]{algorithm2e}

\include{IEEEtran.cls}
\newtheorem{theorem}{Theorem}
\newtheorem{lemma}{Lemma}

\newtheorem{remark}{Remark}

\newtheorem{problem}{Problem}
\newtheorem{example}{Example}
\newtheorem{property}{Property}

\newtheorem{proposition}{Proposition}
\newtheorem{definition}{Definition}
\newcommand{\QED}{{\rm $\blacksquare$}}

\begin{document}

\title{On Non-Orthogonal Multiple Access with Finite-Alphabet Inputs in Z-Channels}
{\author{Zheng Dong, He Chen, Jian-Kang Zhang, and Lei Huang
	\thanks{
		
		Z. Dong is with Shenzhen University, Shenzhen 518060, China, and also with McMaster University, Hamilton, ON L8S 4K1, Canada (e-mail: dongz3@mcmaster.ca).

        H. Chen is with The University of Sydney, Sydney, NSW 2006, Australia (e-mail: he.chen@sydney.edu.au).

       J.-K. Zhang is with McMaster University, Hamilton, ON L8S 4K1, Canada (e-mail: jkzhang@mail.ece.mcmaster.ca).

       L. Huang is with Shenzhen University, Shenzhen 518060, China (e-mail: lhuang@szu.edu.cn).

	}
		}
\maketitle

\begin{abstract}
This paper focuses on the design of non-orthogonal multiple access (NOMA) in a classical two-transmitter two-receiver Z-channel, wherein one transmitter sends information to its intended receiver from the direct link while the other transmitter sends information to both receivers from the direct and cross links. Unlike most existing designs using (continuous) Gaussian input distribution, we consider the practical finite-alphabet (i.e., discrete) inputs by assuming that the widely-used quadrature amplitude modulation (QAM) constellations are adopted by both transmitters. To balance the error performance of two receivers, we apply the max-min fairness design criterion in this paper. More specifically, we propose to jointly optimize the scaling factors at both transmitters, which control the minimum Euclidean distance of transmitting constellations, to maximize the smaller minimum Euclidean distance of two resulting constellations at the receivers, subject to an individual average power constraint at each transmitter. The formulated problem is a \emph{mixed continuous-discrete} optimization problem and is thus intractable in general. By resorting to the Farey sequence, we manage to attain the closed-form expression for the optimal solution to the formulated problem. This is achieved by dividing the overall feasible region of the original optimization problem into a finite number of sub-intervals and deriving the optimal solution in each sub-interval. Through carefully observing the structure of the optimal solutions in all sub-intervals, we obtain compact and closed-form expressions for the optimal solutions to the original problem in three possible scenarios defined by the relative strength of the cross link. Simulation studies are provided to validate our analysis and demonstrate the merits of the proposed design over existing orthogonal or non-orthogonal schemes.


\end{abstract}
%
\IEEEpeerreviewmaketitle
\section{Introduction}
Multiple access technologies have been playing an important role in determining the performance of each generation of mobile communication systems. Based on how the resources are allocated to users, multiple access technologies can generally be categorized into two types: orthogonal multiple access (OMA) and non-orthogonal multiple access (NOMA) ~\cite[Ch.\,6]{Tse05book}. The current generation of cellular networks, known as 4G, and all previous generations have primarily adopted the OMA technologies, which include frequency-division multiple access (FDMA) for 1G, time-division multiple access (TDMA) for 2G, code-division multiple access (CDMA) for 3G, and orthogonal frequency-division multiple access (OFDMA) for 4G~\cite{Dai15}. In these OMA schemes, the resource is partitioned into orthogonal blocks in time/frequency/code domain, and each resource block is then assigned to one single user exclusively. In this sense, there is no inter-user interference in OMA, leading to low-complexity receiver and scheduling algorithms. Moreover, after the resource allocation, the multiple-user problem is divided into several point-to-point problems such that the well-established single-user encoder/decoder techniques can be directly applied. However, early information-theoretic studies showed that compared with NOMA, OMA has lower spectral efficiency as it normally cannot achieve the multi-user channel capacity region~\cite{Vaezi16}. Besides, OMA is not scalable as the total number of orthogonal resources and their granularity strictly limit the maximum number of served users.

Different from OMA, NOMA exploits the power domain to multiplex multiple users together such that they can be served in the same time/frequency/code resources~\cite{Dai15,Ding14spl,Zhiguoding15, Ding16tvt, Vaezi16}. As such, with proper multi-user detection techniques to deal with the inter-user interference at the receiver side (e.g., successive interference cancellation (SIC)~\cite{Saito13}), NOMA is capable of achieving improved spectral efficiency and serving much more users simultaneously. In fact, the uplink and downlink versions of NOMA, well-known as multiple access channel (MAC) and broadcast channel (BC) respectively, have been intensively investigated for several decades in the information theory community, see, e.g.,~\cite{Cover72,Caire03,Wyner74,Cheng93}. However, due to the high complexity of interference cancellation, these studies mainly lied in the theoretical aspects and their results were not implemented in practical communication systems. With the fast advances of hardware, the implementation of NOMA with interference cancellation becomes more affordable and feasible. Actually, NOMA has been regarded as a key enabling technology to meet the unprecedented requirements of 5G wireless networks due to its significant network throughput gain and great potential to support massive connectivity, low latency and user fairness~\cite{Saito13,Benjebbovu13,Saito14,qianli14,liping06vtmag, Dai15,Zhiguoding15,Yuanarxiv16,Andrews14}. Furthermore, a two-user downlink scenario of NOMA, termed multiuser superposition transmission (MUST), has been incorporated in the 3rd Generation Partnership Project (3GPP) Long Term Evolution-Advanced (LTE-A)~\cite{3gpp2015,Lee16msut}.

Most conventional information-theoretic and recent studies on NOMA adopted the assumption of Gaussian input distribution~\cite{Cover72, Caire03, Sun2015cl, Liu2016access,Ding15clrelay,Zhiguoding15,Ding16wclrelay, Ding16jsac, Yuan16tvt}. Although the designs with Gaussian signaling can approach most of the known capacity inner bounds, such as in~\cite{Cover72, Wyner74, Cheng93, Caire03}, their direct implementation in practical communication systems may lead to significant performance loss~\cite{verduit06}.  Moreover, Gaussian signaling will require unaffordable encoding and decoding efforts, which could lead to extremely high hardware cost, huge storage capability, high computational complexity, and long delay. Therefore, Gaussian inputs could arguably be infeasible for current hardware and it acts mostly as the theoretical benchmark. The inputs of practical wireless systems are actually drawn from finite constellations, such as phase shift keying (PSK) modulation or quadrature amplitude modulation (QAM), which are essentially different from the continuous Gaussian inputs.  When it comes to a NOMA system with finite input constellations, the key design challenge is to guarantee that each user's codeword can be uniquely decoded from their sum signal at the receiver side~\cite{Shu76, Shu78,Ahlswede99}. For the two-user MAC with finite-alphabet inputs, a constellation rotation (CR) scheme and a constellation power allocation (CPA) scheme were proposed in～\cite{Harshan11} and～\cite{Rajan13} to construct an unambiguous sum constellation at the receiver, respectively. This is achieved by strategically introducing certain angle of rotation between the input constellations in the CR scheme and appropriately controlling the transmit power of each user in the CPA scheme. The results in～\cite{Harshan11} and～\cite{Rajan13} have been extended to various multiple-antenna scenarios, see, e.g.,~\cite{Xiaotc15} and references therein. The aforementioned NOMA designs were primarily for the PSK modulations by utilizing its circular structure. The studies on NOMA with QAM, another practical modulation scheme that has been widely adopted in cellular systems due to its higher spectral efficiency, are quite limited. Very recently, the mutual information were used as the performance metric to optimize NOMA systems with QAM in \cite{Choi2016cl,Shieh2016tcom}. However, the optimal NOMA designs in~\cite{Choi2016cl,Shieh2016tcom} were achieved by numerical approaches with high computational complexities.

We also notice that the existing studies on NOMA mainly focused on MAC and BC, due to their wide applications in centralized systems like cellular networks. Also there are some initial efforts considering the NOMA design for the interference channel (IC)~\cite{Vaezi16}.  With recent advances in non-centralized networks (e.g., wireless sensor networks and ad hoc networks), Z-channel (ZC) was proposed in~\cite{Sriram03} and attracted considerable attention in the past decade~\cite{Chong07,Ulukus04,Jafar09isit,Salehkalaibar10itw,Lee13vtc,Liang13itw,Liu14tit}. As a special case of the classical two-user IC~\cite{Kobayashi81}, a two-user ZC consists of two transmitters and two receivers, wherein one transmitter sends information to its intended receiver from the direct link without causing interference to the unintended receiver, while the other transmitter sends information to both receivers from the direct and cross links. In this sense, the ZC is a general channel model that includes MAC and BC as special cases. The ZC is also closely related to the Z-interference channel (ZIC)~\cite{Costa85, Sun15, Kolte14isit, Vaezi16icc}, wherein each transmitter transmits information to its corresponding receiver only from the direct link and the received signal from the  cross link carries no desired information, and thus is treated as interference at the receiver side. It is also worth emphasizing that there are two messages transmitted from the two direct links in ZIC, while three messages are sent via the two direct links and the cross link in ZC.
Both ZC and ZIC are proper models for the multi-cell downlink transmission, where one user is located near to the cell edge and thus can receive signals from both transmitters, while the other user is near the cell center and suffers from no interference. Another example corresponds to the two-user IC, where one cross link is blocked by obstacles with large pathloss such as tall buildings or thick walls, while the other user is still exposed to interference~\cite{Sun15, Vidal16}. Despite of the existing great efforts on ZC, to our best knowledge, the design of NOMA with practical finite-alphabet inputs in ZC is still an \emph{open problem} in the literature.

Motivated by this gap, in this paper we concentrate on the practical design of NOMA with QAM in a two-transmitter two-receiver ZC. It is worth emphasizing that the design of NOMA with QAM is much more challenging than that for PSK modulation. This is mainly because the unambiguity of sum QAM constellations is much harder to maintain since its signal points are distributed more evenly and there is a higher probability that more than one signal points coincide or close to each other on the sum constellation. The main contributions of this paper can be summarized as follows:
\begin{enumerate}
	\item We, for the first time, develop a practical NOMA framework with QAM and max-min fairness in ZC. In our framework, we optimize the scaling factors of both transmitters, which adjust the minimum Euclidean distance of the transmitting constellations, to maximize the smaller minimum Euclidean distance among the resulting constellations at both receivers subject to an individual power constraint on each transmitter. Through our design, the average error performance of both transmitters in the considered ZC can be minimized with good user fairness, which is fundamentally different from the existing designs that mainly focused on the channel capacity maximization.
	\item The formulated optimization problem is shown as a mixed continuous-discrete optimization problem, which is challenging to solve in general. By carefully observing the features of the formulated problem, we realize that the Farey sequence (also known as Farey series)~\cite{Hardy75} can be applied to resolve the problem. More specifically, by taking the advantage of Farey sequence associated with the finite-alphabet, we strategically partition the entire feasible region of the original optimization problem into a finite number of sub-intervals and attain the closed-form solution in each sub-interval. Then, by a careful observation on the structure of the solutions in all sub-intervals, the overall solution is obtained in a compact closed-form for three complementary scenarios divided by the relative strength of the cross link.
	\item We verify the correctness of the analytical results by conducting simulations in both deterministic and random fading channels.
Simulation results show that the sum constellation at the receiver side is still a regular QAM constellation with a larger size for most scenarios, but could be a hierarchical QAM with two-resolution~\cite{Vetterli93}, e.g., when the cross link is very strong relative to the direct link.
We adopt the bit error rate (BER) as the performance metric to compare the proposed NOMA design with the existing OMA and NOMA schemes under random fading channels. The comparison illustrates that our scheme can achieve a significant lower BER performance than the benchmark schemes, which validates the effectiveness of our design.
\end{enumerate}	


\section{System Model of Complex Gaussian ZC with QAM Constellations}
\begin{figure}	\centering
	\subfigure[]{\label{fig:twousric}
		\includegraphics[width=0.45\linewidth]{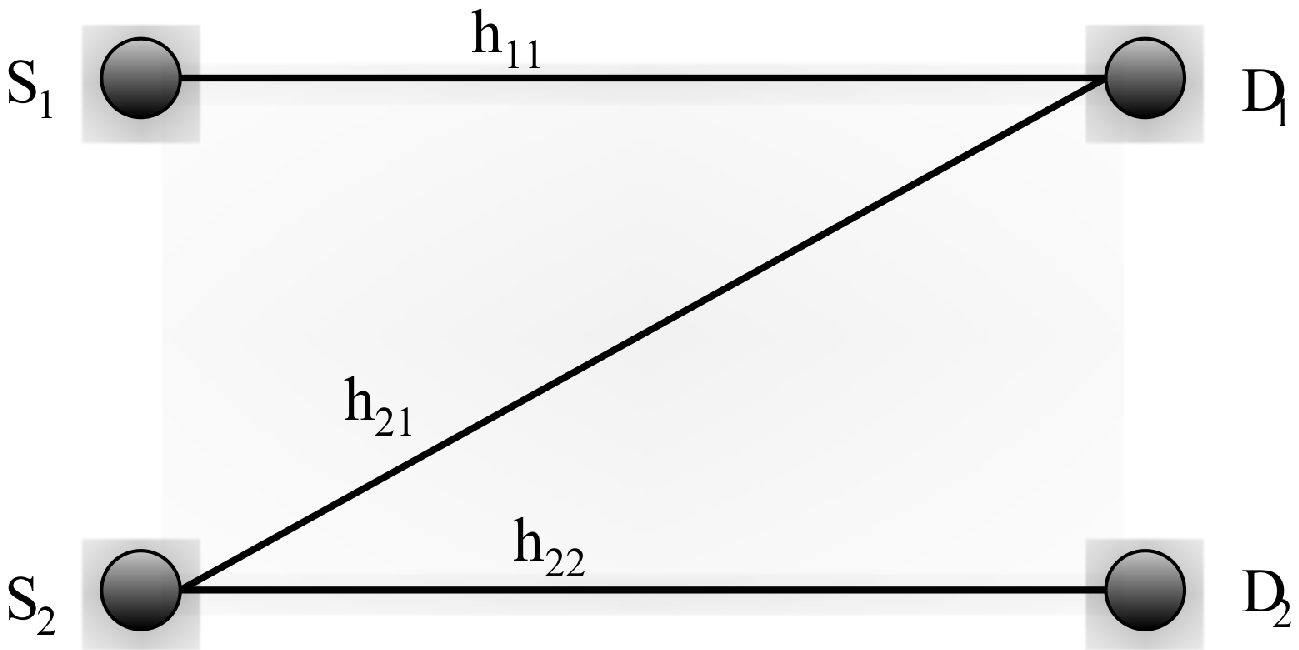}
		 }
	\subfigure[]{\label{fig:gaussianzchannel}
		\includegraphics[width=0.45\linewidth]{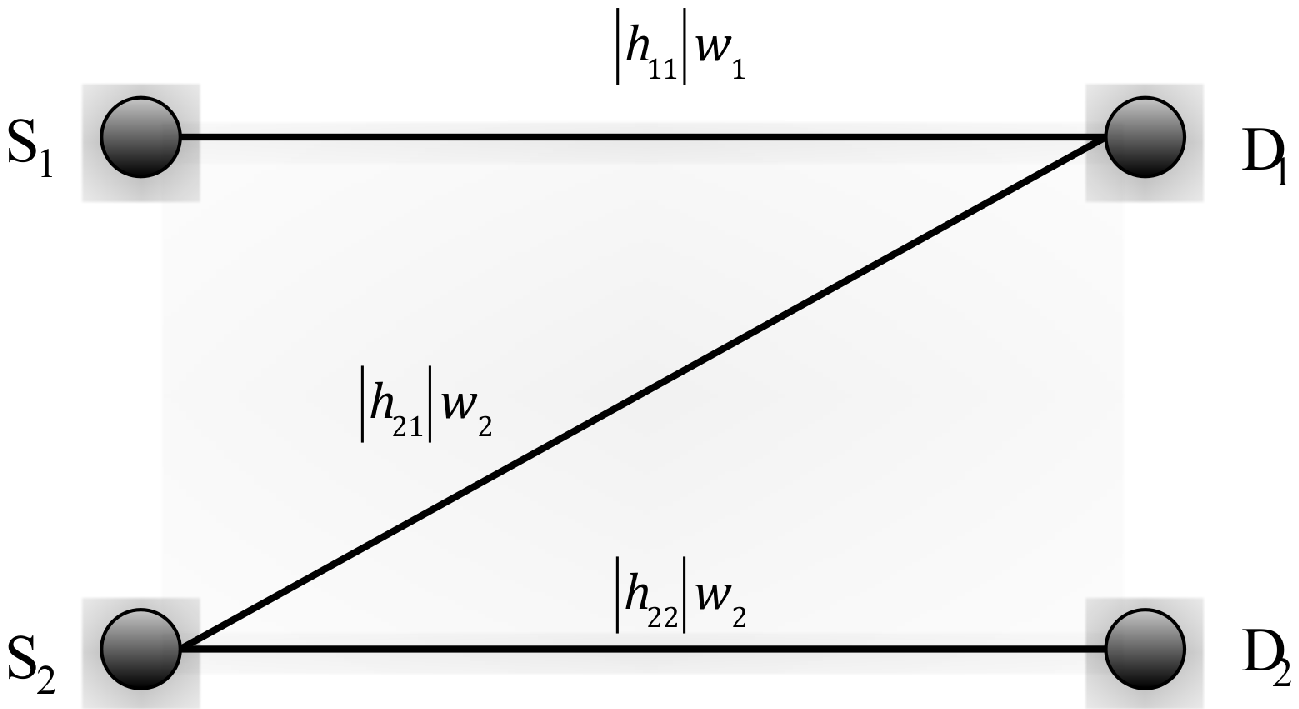}
		}
	\vspace{-10pt} \caption{(a) Two-user complex Gaussian ZC.~(b) Two-user real Gaussian ZC.}
\end{figure}
We consider a two-user complex Gaussian ZC consisting of two transmitters $S_1$ and $S_2$, and two receivers $D_1$ and $D_2$, as depicted in Fig.\,\ref{fig:twousric}. We consider that each node is equipped with a single antenna and works in a half-duplex mode. As per the ZC, only the cross link between $S_2$ and $D_1$ is assumed to be available.
Moreover, $S_1$ sends one unicasting message $x_1$ to $D_1$, while $S_2$ transmits one multicasting message $x_2$ to both $D_1$ and $D_2$. $S_1$ and $S_2$ transmit their messages simultaneously using the same frequency band. The equivalent complex baseband signals observed at $D_1$ and $D_2$ can be given, respectively, by the following equations:
\begin{subequations}\label{eqn:comlexZC}
	\begin{align}
	z_1 &=h_{11} x_1+ h_{21} x_2 +\xi_1,\\
	z_2 &=h_{22} x_2 +\xi_2,
	\end{align}
\end{subequations}
where $h_{k\ell} \in \mathbb C$, $k,\ell\in\{1,2\}$ denotes the complex channel coefficients from transmitter $S_k$ to receiver $D_\ell$. Hereafter, we call $h_{11}$ and $h_{22}$  the direct links, while $h_{21}$ is referred to as the cross link. In line with~\cite{Sriram03,Chong07, Ulukus04,Jafar09isit,Salehkalaibar10itw,Liang13itw,Liu14tit,Sun15}, all the channel links  are assumed to be known perfectly at all the terminals. The additive noise processes $\xi_1, \xi_2 \sim \mathcal{CN}(0, 2\sigma^2)$ are independent and identically distributed (i.i.d.) over time and are assumed to be circularly symmetric complex Gaussian (CSCG). Note that the case with different noise levels at receivers can be incorporated into our model by scaling operations. We suppose that QAM constellations are used by both transmitters since it is more spectrally efficient than other frequently-used modulation schemes such as phase-shift keying (PSK), and is also relatively easy to implement~\cite[Ch.\,5.3.3]{goldsmith05}. A predefined average transmitted power constraint is imposed to both transmitters\footnote{Our design can also be generalized to the case with peak power constraint straightforwardly.}, i.e., $\mathbb E[|x_1|^2]\le P_1$ and $\mathbb E[|x_2|^2] \le P_2$. In this paper, the system signal-to-noise ratio (SNR) is defined by $\rho=\frac{1}{2\sigma^2}$.

For analytical simplicity, we decompose the considered complex Gaussian ZC given in~\eqref{eqn:comlexZC} into two parallel real scalar Gaussian ZCs~\cite{Sriram03,Chong07,Sun15}, which are called the in-phase and quadrature components, respectively. We note that this method was commonly used in the study of IC and the real IC was studied directly in~\cite{Cadambe08,Tse08, Bresler10, Vidal16}.
Actually, designing two-dimensional QAM constellations is an extremely challenging problem even for two-user MAC, see e.g.,~\cite{Harshan11, Rajan13} and references therein. In this paper, instead of designing the two-dimensional QAM constellations directly, we propose a practical design that decomposes the complex Gaussian ZC into two parallel real scalar Gaussian ZCs such that we can split the two-dimensional QAM constellation into two one-dimensional PAM constellations. In fact, although we use complex baseband representation in~\eqref{eqn:comlexZC}, the actual modulated and demodulated signals are all real since the oscillator at the transmitter can only generate real sinusoids rather than complex exponentials, and the channel just introduces amplitude and phase distortion to the transmitted signals~\cite[Ch.\,2.2]{goldsmith05}. By this means, the original two-dimensional QAM constellation can be split into two one-dimensional pulse amplitude modulation (PAM) constellations for both the in-phase and quadrature components.  Mathematically, for the complex Gaussian ZC described in~\eqref{eqn:comlexZC}, the in-phase and quadrature components can be attained by rotating $x_1$ and $x_2$ according to the instantaneous channel coefficients to compensate for the phase offset, and then taking the real and imaginary parts, respectively. First of all, we note that~\eqref{eqn:comlexZC} is equivalent to
	\begin{align}
	&z_1 =|h_{11}|\exp(j\arg(h_{11})) x_1+|h_{21}|\exp(j\arg(h_{21})) x_2+\xi_1 \nonumber\\
	&\exp \Big(j \arg\big( \frac{h_{21}}{h_{22}}\big)\Big)z_2=|h_{22}| \exp(j\arg(h_{21})) x_2 \nonumber\\
	&\qquad\qquad\qquad\qquad\qquad\qquad+\exp \Big(j \arg\big( \frac{h_{21}}{h_{22}}\big)\Big)\xi_2.
	\end{align}

Now, we set
\begin{subequations}\label{eqn:complex2real}
\begin{align}
&y_1={\rm Re}(z_1),~y_2={\rm Re}\Big(\exp (j \arg \big( \frac{h_{21}}{h_{22}}\big))z_2\Big),\\
&w_1 s_1={\rm Re} \big(\exp(j\arg(h_{11}))x_1\big),\\
&w_2 s_2={\rm Re} \big(\exp (j \arg(h_{21}))x_2 \big),\\
&n_1={\rm Re}(\xi_1), ~ n_2= {\rm Re} \big(\exp(j \arg\big( \frac{h_{21}}{h_{22}}\big)) \xi_2\big); \\
&y_1'={\rm Im}(z_1), ~ y_2'={\rm Im}\Big(\exp (j \arg\big( \frac{h_{21}}{h_{22}}\big))z_2\Big),\\
&w_1' s_1'={\rm Im} \big(\exp(j\arg(h_{11}))x_1\big), \\
&w_2' s_2'={\rm Im}\big(\exp (j \arg(h_{21}))x_2 \big),\\
&n_1'={\rm Im}(\xi_1), ~ n_2'={\rm Im}\big(\exp(j \arg\big( \frac{h_{21}}{h_{22}}\big)) \xi_2\big),
\end{align}
\end{subequations}
where ${\rm Re}(\cdot)$ and ${\rm Im}(\cdot)$ are the real and  imaginary parts of the complex number, respectively.
We also assume that $s_1 \in \mathcal{A}_{M_1}$, $s_1' \in \mathcal{A}_{M_1'}$, sent by $S_1$, and $s_2 \in \mathcal{A}_{M_2}$,  $s_2' \in \mathcal{A}_{M_2'}$, transmitted by $S_2$, are the information-bearing symbols, which are drawn from standard PAM constellation with equal probability, in which $\mathcal{A}_{M}\ \triangleq\{\pm (2k-1)\}_{k=1}^{{M}/2}$ is a $M$-ary PAM constellation set. Moreover, the scaling factors $w_1$, $w_2$, $w_1'$, and $w_2'$ are real positive scalars that determine the minimum Euclidean distance of the corresponding PAM constellation set.

From~\eqref{eqn:complex2real}, we obtain $\exp\big(j\arg(h_{11})\big) x_1=  (w_1 s_1 + w_1' s_1'j) \in \mathcal{Q}_1$ and $\exp \big(j\arg(h_{21})\big) x_2= (w_2 s_2 + w_2' s_2'j) \in \mathcal{Q}_2$, where
\begin{subequations}
	\begin{align}
	&\mathcal{Q}_1 \triangleq \{\pm w_1 (2k-1) \pm w_1' (2\ell-1)j: \nonumber\\
	&\qquad\qquad\qquad k=1,\ldots, M_1/2, \ell=1,\ldots, M_1'/2\},\\
    &\mathcal{Q}_2 \triangleq  \{\pm w_2 (2k-1) \pm w_2' (2\ell-1)j:\nonumber\\
    &\qquad\qquad\qquad k=1,\ldots, M_2/2, \ell=1,\ldots, M_2'/2\},
	\end{align}
\end{subequations}	
are $M_1 M_1'$- and $M_2 M_2'$-ary QAM constellations, respectively.
If $w_1=w_1'$ and $w_2=w_2'$, we call $\mathcal{Q}_1$ and $\mathcal{Q}_2$ symmetric QAM constellations. Otherwise, we are using unsymmetric signaling~\cite{Sun15,Vidal16, Jafar10}. In addition, $n_1, n_2, n_1', n_2'\sim \mathcal{N}(0, \sigma^2)$ are i.i.d. real additive white Gaussian since the complex noise terms are assumed to be CSCG.
Then, the in-phase and quadrature sub-channels of~\eqref{eqn:comlexZC}, as illustrated in Fig.~\ref{fig:gaussianzchannel}, can be reformulated by
\begin{subequations}\label{eqn:realZC}
\begin{align}
y_1 &= |h_{11}|w_1 s_1 + |h_{21}| w_2 s_2 +n_1, \nonumber\\
y_2 &= |h_{22}|w_2 s_2 +n_2,\label{eqn:inphaseZC}\\	
y_1'&= |h_{11}|w_1' s_1' + |h_{21}| w_2' s_2' +n_1',\nonumber\\
y_2'&= |h_{22}|w_2' s_2' +n_2'.
\end{align}
\end{subequations}
The transmitted signals over both subchannels should still be subject to average power constraints, i.e., $\mathbb E[w_1^2 |s_1|^2]\le p_1$, $\mathbb E[w_2^2|s_2|^2]\le p_2$,  $\mathbb E[w_1'^2 |s_1'|^2]\le p_1'$, $\mathbb E[w_2'^2|s_2'|^2]\le p_2'$ such that $p_1+p_1'=P_1$ and $p_2+p_2'=P_2$.
The following power allocation among the in-phase and quadrature components is normally performed to balance the minimum Euclidean distance of the two PAM constellations~\cite[Ch.\,6.1.4]{goldsmith05}, i.e.,
\begin{subequations}\label{eqn:powerallocation}
	\begin{align}
	p_1&=\frac{(M_1^2-1)P_1}{M_1^2+M_1'^2-2},~
	p_2=\frac{(M_2^2-1)P_2}{M_2^2+M_2'^2-2},\\
	p_1'&=\frac{(M_1'^2-1)P_1}{M_1^2+M_1'^2-2},~
	p_2'=\frac{({M}_2'^2-1)P_2}{M_2^2+M_2'^2-2}.
	\end{align}
\end{subequations}
It can be observed that, if square-QAM constellations are used at both transmitters with $M_1=M_1'$ and $M_2=M_2'$, we have $p_1=p_1'=P_1/2$ and $p_2=p_2'=P_2/2$.

An important problem for the considered ZC is that for any given QAM constellation sizes of both messages, how to optimize the values of scaling coefficients $w_1$, $w_2$, $w_1'$ and $w_2'$ to minimize the average error probability at both receivers, subject to the individual average power constraint at both transmitters. By leveraging the decomposable property of the complex Gaussian ZC and the symmetry of the two subchannels, we can simply focus on the design for one of the two real Gaussian ZCs with PAM constellation sets, which will be elaborated in the next section\footnote{It should be pointed out that this design is a practical but not necessarily optimal approach, which has been widely adopted in practice~\cite{Chong07,Cadambe08,Tse08, Jafar09,Jafar10,Bresler10, Sun15, Prasad16, Vidal16}.}.


\section{The Constellation Design for the Real Gaussian ZC}
In this section, we consider the constellation design problem, i.e., finding the optimal values of $w_1$ and $w_2$ for the in-phase real Gaussian ZC characterized by~\eqref{eqn:inphaseZC}. The optimal solution to the  quadrature component can be obtained in a similar fashion and hence omitted for brevity. In particular, if $M_1=M_1'$ and $M_2=M_2'$, then the two sub-channels are identical.  It is worth noting that, similar design for BC or MAC can be included as a special case of our proposed design for the considered ZC.

\subsection{Problem Formulation}
As the first effort towards the design of NOMA with finite-alphabet inputs in ZC, in this paper we concentrate on the case that $M_1=M_2=M$ and $M_1'=M_2'=M'$.
As a result, we have $s_1, s_2 \in \mathcal{A}_{M}=\{\pm (2k-1)\}_{k=1}^{{M}/2}$. As $\mathbb E[w_1^2 |s_1|^2]\le p_1$ and $\mathbb E[w_2^2|s_2|^2]\le p_2$, we thus have  $0< w_1 \le \sqrt{\frac{3 p_1}{M^2-1}}$ and $0< w_2 \le \sqrt{\frac{3 p_2}{M^2-1}}$.

In our scheme, the transmitted signal from $S_1$ and $S_2$ are superimposed together at $D_1$, which is inherently a non-orthogonal transmission. In line with~\cite{Chong07}, we use a \emph{joint decoding}\footnote{We note that, for ZIC, the joint decoder used by $D_1$ may not necessarily be the most efficient one. Instead, we should use a treat-interference-as-noise (TIN) receiver when the channel gain of the cross link is very low and use a successive-interference-cancellation (SIC) receiver when the channel gain of cross link is very strong compared with the direct link~\cite{Costa85,Chong07,Kobayashi81}. In general, a joint decoder can be used when the cross link is moderately strong~\cite{Kobayashi81,Chong07}, which will result in a similar design as our case. However, how to extend our design to ZIC with finite-alphabet input is still an open problem and has been left as a future work.}
} at the receiver $D_1$ since the error performance is dominated by the minimum Euclidean of the resulting sum-constellation at $D_1$.
We assume that each receiver uses a coherent maximum-likelihood~(ML) detector to estimate the transmitted signals in a symbol-by-symbol fashion\footnote{Since we are doing a symbol-by-symbol detection, the decoding complexity is $\mathcal{O}(M^2)$ with $M$ being the PAM constellation size of $S_1$ and $S_2$, respectively. Although we can use the message passing algorithm (MPA)~\cite{Hoshyar08tsp} to further decrease the decoding complexity, however, our method is feasible.}.
For receivers $D_1$ and $D_2$, the estimated signals can be expressed as
\begin{align*}
&(\hat s_1,\hat s_2)=\arg \min_{(s_1, s_2)}~\big|y_1 -(|h_{11}|w_1 s_1 + |h_{21}| w_2 s_2)\big|,\\
& \hat s_2' =\arg \min_{s_2}~\big|y_2 -|h_{22}|w_2 s_2\big|.
\end{align*}

By applying the nearest neighbor approximation method~\cite[Ch.\,6.1.4]{goldsmith05} at high SNRs for the ML receiver, the average error rate is dominated by the minimum Euclidean distance of the received constellation points owing to the exponential decaying of the Gaussian distribution. To balance the error performance of both receivers, in this paper, we aim to devise the optimal value of $w_1$ and $w_2$ by applying the max-min fairness criterion on the minimum Euclidean distance of the received signal constellation points among both receivers.

The Euclidean distance between two received signals $y_1(s_1,s_2)$ and $y_1(\tilde{s}_1, \tilde{s}_2)$ at $D_1$ and that between $y_2(s_2)$ and $y_2(\tilde{s}_2)$ at $D_2$
 for the transmitted signal vectors $(s_1, s_2)$ and $(\tilde{s}_1, \tilde{s}_2)$ at $S_1$ and $S_2$ in the noise-free case are given, respectively, by
\begin{subequations}
	\begin{align}
	&|y_1(s_1, s_2) -y_1(\tilde{s}_1, \tilde{s}_2)|=\big||h_{11}| w_1(s_1 -\tilde{s}_1)\nonumber \\
	&\qquad\qquad\qquad\qquad\qquad\qquad  -|h_{21}| w_2 (\tilde{s}_2-s_2 ) \big|,\\
	&|y_2(s_2) -y_2(\tilde{s}_2)|=|h_{22}|w_2 |s_2 -\tilde{s}_2|.
	\end{align}
\end{subequations}

Note that $s_1$, $\tilde{s}_1$, $s_2$ and $\tilde{s}_2$ are all odd numbers, and we thus can let $s_1 -\tilde{s}_1=2 n$ and $\tilde{s}_2-s_2 =2 m$, in which $m, n\in \mathbb Z_{M-1}$ with $\mathbb Z_{M-1}  \triangleq \{0,\pm 1,\ldots, \pm M-1\}$  denoting the set containing all the possible differences.  Similarly, we also define $\mathbb Z_{M-1}^2\triangleq \{(a,b):a,b \in \mathbb Z_{M-1}\}$, $\mathbb {N}_{M-1}\triangleq \{0,1,\cdots,M-1\}$ and $\mathbb N_{M-1}^2 \triangleq \{(a,b): a,b \in \mathbb N_{M-1}\}$. From the above definition, $(s_1, s_2) \neq (\tilde{s}_1, \tilde{s}_2)$ is equivalent to $(m,n) \neq (0,0)$. Here, by $(m,n) \neq (0,0)$, we mean that $m \neq 0$ or $n \neq 0$. To proceed, we define
\begin{subequations}\label{eqn:euclideandist}
\begin{align*}
&d_1(m,n)=\frac{1}{2}|y_1(s_1, s_2) -y_1(\tilde{s}_1, \tilde{s}_2)|\\
&\quad=\big||h_{11}|w_1 n -  |h_{21}| w_2 m\big|, {\rm for}~(m,n) \in \mathbb Z^2_{M-1} \setminus\{(0,0)\},\\
&d_2(m)=\frac{1}{2}|y_2(s_2) -y_2(\tilde{s}_2)|\\
&\quad=|h_{22}| w_2 |m|, \quad {\rm for}~m \in \mathbb Z_{M-1} \setminus \{0\}.
\end{align*}
\end{subequations}

We are now ready to formally formulate the following max-min optimization problem:

\begin{problem}[Optimal Design of NOMA in real scalar ZC with PAM constellation]\label{pbm:maxminopt}
Find the optimal value of $(w_1^*,w_2^*)$ subject to the individual average power constraint such that the minimum value of the minimum Euclidean distances of the received signal constellation points over both received signals is maximized, i.e.,
	\begin{subequations}
	\begin{align}
	&(w_1^*, w_2^*)=\arg\max_{(w_1, w_2)}\,\min \Big\{\underbrace{\min_{ (m,n) \in \mathbb Z_{M-1}^2 \setminus\{(0,0)\} } d_1(m,n)}_{T_1},\nonumber\\
	&\qquad\qquad\qquad\qquad\qquad\qquad\quad\underbrace{ \min_{m \in \mathbb Z_{M-1} \setminus \{0\}} d_2(m)}_{T_2}  \Big\} \label{eqn:maxminoptobj}\\
	&{\rm s.t.~} 0< w_1 \le \sqrt{\frac{3 p_1}{M^2-1}}{~\rm~and~} 0< w_2 \le \sqrt{\frac{3 p_2}{M^2-1}}.
	\end{align}
	\end{subequations}
	\hfill$\blacksquare$
\end{problem}

Note that the inner optimization problem of finding the minimum Euclidean distances is a discrete one, while the outer optimization problem on $(w_1, w_2)$ is a continuous problem. In other words, Problem~\ref{pbm:maxminopt} is a \emph{mixed continuous-discrete} optimization problem and it is in general hard to solve. To the best of our knowledge, only numerical solutions to such kind of problems are available in the literature, see e.g.,~\cite{Harshan11,Rajan13,Xiaotc15} and references therein.

To optimally and systematically solve this problem, we now develop a novel framework based on the \emph{Farey sequence} (also known as Farey series)~\cite{Hardy75}, which can divide the entire feasible region of $(w_1, w_2)$ into a finite number of mutually exclusive sub-intervals.
Then for each sub-interval, the formulated optimization problem can be solved optimally with a closed-form solution, and subsequently the overall maximum value of Problem~\ref{pbm:maxminopt} can be attained by taking the maximum value of the objective function among all the sub-intervals.

For the inner optimization problem of $T_2$ given in~\eqref{eqn:maxminoptobj}, it can be observed that
\begin{multline*}
\min_{m \in \mathbb Z_{M-1} \setminus \{0\}}~d_2(m)=\min_{m \in \mathbb Z_{M-1} \setminus \{0\}} |h_{22}| w_2|m|\\
\qquad=|h_{22}| w_2,~{\rm with~}m=1.
\end{multline*}
However, for $T_1$, we have
\begin{multline}\label{eqn:mindistD1}
\min_{ (m,n)\in \mathbb Z_{M-1}^2 \setminus\{(0,0)\} }~d_1(m,n) \\
\qquad= \min_{\quad (m,n)\in \mathbb Z_{M-1}^2 \setminus\{(0,0)\}} \big||h_{11}|w_1 n -  |h_{21}| w_2 m\big|.
\end{multline}
We should point out that the closed-form solution to the optimal $(m,n)$ is not trivial, since the solution depends on the values of $|h_{11}|$ and $|h_{22}|$, which can span the whole positive real axis. Moreover, the value of $w_1$ and $w_2$ can not be determined beforehand. Actually, the problem in~\eqref{eqn:mindistD1}  is essentially  equivalent to finding a real rational number with finite order to approximate a real irrational number as closely as possible. This naturally leads us to resorting to the Farey sequence, which particularly plays a critical role in solving such kind of problems~\cite{Hardy75}. In the subsequent section, we will introduce the definition and some important properties of Farey sequence.

\subsection{Farey Sequence}
The Farey sequence characterizes the relationship between two positive integers and the formal definition is  given as follows:
\begin{definition}[Farey sequence]\cite{Hardy75}
The Farey sequence $\mathfrak{F}_K$ of order $K$ is the ascending sequence of irreducible fractions between $0$ and $1$ whose denominators are less than or equal to $K$.
\hfill\QED
\end{definition}	

By the definition, $\mathfrak{F}_K=\big(\frac{b_k}{a_k}\big)_{k=1}^{|\mathfrak{F}_K|}$ is a sequence of fractions $\frac{b_k}{a_k}$ such that $0\le b_k\le a_k\le K$ and $\langle a_k, b_k \rangle =1$  arranged in an increasing order, where $\langle a,b\rangle$ denotes the largest common divider of non-negative integers $a,b$. $|\mathfrak{F}_K| =1+\sum_{m=1}^K \varphi(m)$ is the cardinality of $\mathfrak{F}_K$ with $\varphi(\cdot)$ being the Euler's totient function~\cite{Hardy75}.  Some examples of Farey sequences are given as follows:
\begin{example}
    $\mathfrak{F}_5$ is the ordered sequence
	\begin{align*}
	\Big(\frac{0}{1}, \frac{1}{5}, \frac{1}{4}, \frac{1}{3}, \frac{2}{5},\frac{1}{2}, \frac{3}{5}, \frac{2}{3},  \frac{3}{4}, \frac{4}{5}, \frac{1}{1}\Big).
	\end{align*}
\end{example}

It can be observed that each Farey sequence begins with number 0 (fraction $\frac{0}{1}$) and ends with 1 (fraction $\frac{1}{1}$). The series of breakpoints after $\frac{1}{1}$ is the reciprocal version of the Farey sequence. We call the Farey number sequence together with its reciprocal version as the \emph{extended Farey sequence} which is formally defined as follows:
\begin{definition}[Extended Farey sequence]
The extended Farey sequence $\mathfrak{S}_K$ of order $K$ is the sequence of ascending irreducible fractions, where the maximum value of the numerator and denominator do not exceed $K$.
\hfill\QED
\end{definition}
	
From the definition, we have $\mathfrak{S}_K=\big(\frac{b_k}{a_k}\big)_{k=1}^{|\mathfrak{S}_K|}$ with $\langle a_k,b_k\rangle=1$ and  $|\mathfrak{S}_K|=1+2\sum_{m=1}^K \varphi(m)$. We have the following example:
\begin{example}
	$\mathfrak{S}_5$ is the sequence
	\begin{align*}
	\Big(\frac{0}{1}, \frac{1}{5}, \frac{1}{4}, \frac{1}{3}, \frac{2}{5},\frac{1}{2}, \frac{3}{5}, \frac{2}{3},  \frac{3}{4}, \frac{4}{5}, \frac{1}{1},\frac{5}{4},\frac{4}{3},\frac{3}{2},\frac{5}{3},\frac{2}{1},\frac{5}{2},\frac{3}{1},\frac{4}{1},\frac{5}{1},\frac{1}{0}\Big).
	\end{align*}
\end{example}
It can be observed that the extended Farey sequence starts with number 0 (fraction $\frac{0}{1}$) and end with $\infty$ (fraction $\frac{1}{0}$).

The positive real axis can be divided by the extended Farey sequence $\mathfrak{S}_K$ into a \emph{finite number} (i.e., $|\mathfrak{S}_K|-1=2\sum_{m=1}^K \varphi(m)$)  of intervals. In this paper, we call the fractions consisting of adjacent terms in the extended Farey sequence as a \emph{Farey pair}, and the interval between the Farey pair is referred to as a \emph{Farey interval}. We then have the Farey interval set formally defined as follows:
\begin{definition}[Farey interval set]
A Farey interval set $\mathcal{S}_K$ of order $K$ is the set containing all the Farey intervals generated by the Farey pair of the extended Farey sequence $\mathfrak{S}_K$.\hfill\QED	
\end{definition}	

By the definition, we have $\mathcal{S}_K=\big\{\big(\frac{b_k}{a_k}, \frac{b_{k+1}}{a_{k+1}}\big)\big\}_{k=1}^{|\mathcal{S}_K|}$, where $|\mathcal{S}_K|=|\mathfrak{S}_K|-1=2\sum_{m=1}^K \varphi(m)$. Note that,  with a slight abuse of notations, $\big(\frac{b_k}{a_k}, \frac{b_{k+1}}{a_{k+1}}\big)$ denotes the interval between end nodes $\frac{b_k}{a_k}$ and $\frac{b_{k+1}}{a_{k+1}}$ rather than a vector, and this will be clear from the context.
\begin{example}
The	Farey interval set $\mathcal{S}_5$ is the set given by
	\begin{align*}
	\Big\{\Big(\frac{0}{1}, \frac{1}{5}\Big), \Big(\frac{1}{5}, \frac{1}{4}\Big), \Big(\frac{1}{4}, \frac{1}{3}\Big), \ldots, \Big(\frac{3}{1}, \frac{4}{1}\Big), \Big(\frac{4}{1}, \frac{5}{1}\Big),\Big(\frac{5}{1}, \frac{1}{0}\Big)\Big\}.
	\end{align*}
\end{example}

The Farey interval set can be further divided into two subsets $\mathcal{U}_K^L=\{\big(\frac{b_k'}{a_k'}, \frac{b_{k+1}'}{a_{k+1}'}\big)\in \mathcal{S}_K: b_k' +b_{k+1}' \ge L\}$  and $\mathcal{V}_K^L=\{\big(\frac{b_k''}{a_k''}, \frac{b_{k+1}''}{a_{k+1}''}\big)\in \mathcal{S}_K: b_k'' +b_{k+1}'' < L\}$ for $L=1,2,\ldots, 2K$ such that $\mathcal{S}_K=\mathcal{U}_K^L \cup \mathcal{V}_K^L$ and $\mathcal{U}_K^L \cap \mathcal{V}_K^L=\varnothing$. In particular, $\mathcal{U}_K^1 = \mathcal{S}_K$ and $\mathcal{V}_K^1=\varnothing$ while $\mathcal{U}_K^{2K}=\varnothing$ and $\mathcal{V}_K^{2K}=\mathcal{S}_K$.

\begin{example}\label{exp:subseq}
For the	Farey interval set $\mathcal{S}_5$, we have
\begin{align*}
\mathcal{U}_5^4=&\Big\{
\Big(\frac{1}{2}, \frac{3}{5}\Big),
\Big(\frac{3}{5}, \frac{2}{3}\Big),
\Big(\frac{2}{3}, \frac{3}{4}\Big),
\Big(\frac{3}{4}, \frac{4}{5}\Big),
\Big(\frac{4}{5}, \frac{1}{1}\Big),\\
&\quad\Big(\frac{1}{1}, \frac{5}{4}\Big),
\Big(\frac{5}{4}, \frac{4}{3}\Big),
\Big(\frac{4}{3}, \frac{3}{2}\Big),
\Big(\frac{3}{2}, \frac{5}{3}\Big),
\Big(\frac{5}{3}, \frac{2}{1}\Big),\\
&\qquad\Big(\frac{2}{1}, \frac{5}{2}\Big),
\Big(\frac{5}{2}, \frac{3}{1}\Big),
\Big(\frac{3}{1}, \frac{4}{1}\Big),
\Big(\frac{4}{1}, \frac{5}{1}\Big),
\Big(\frac{5}{1}, \frac{1}{0}\Big)
\Big\},
\end{align*}
and
\begin{align*}
\mathcal{V}_5^4=&\Big\{
\Big(\frac{0}{1}, \frac{1}{5}\Big),
\Big(\frac{1}{5}, \frac{1}{4}\Big),
\Big(\frac{1}{4}, \frac{1}{3}\Big),
\Big(\frac{1}{3}, \frac{2}{5}\Big),
\Big(\frac{2}{5}, \frac{1}{2}\Big)
\Big\}.
\end{align*}
\end{example}

We now review some elementary properties of Farey sequences~\cite{Hardy75} which are also true for extended Farey sequences.
\begin{property}\label{prop:basicFareyprop}
If $\frac{n_1}{m_1}$ and $\frac{n_2}{m_2}$ are  two adjacent terms (Farey pairs) of $\mathfrak{S}_K$ with $K\ge 1$, such that $\frac{n_1}{m_1}< \frac{n_2}{m_2}$, then,
\begin{enumerate}
	\item $m_1 n_2 - m_2 n_1=1$.
	\item $\frac{n_1 + n_2}{m_1 + m_2} \in \big(\frac{n_1}{m_1}, \frac{n_2}{m_2} \big)$, $\frac{m_1+m_2}{n_1 + n_2} \in \big(\frac{m_2}{n_2}, \frac{m_1}{n_1} \big)$.~\hfill$\blacksquare$
\end{enumerate}

\end{property}

\begin{property}
	If $\frac{n_1}{m_1}$, $\frac{n_2}{m_2}$  and $\frac{n_3}{m_3}$ are three consecutive terms of $\mathfrak{S}_K$ with $K\ge 1$ such that $\frac{n_1}{m_1}< \frac{n_2}{m_2}<\frac{n_3}{m_3}$, then, $\frac{n_2}{m_2}=\frac{n_1 + n_3}{m_1 + m_3}$.
	~\hfill$\blacksquare$
\end{property}

\begin{property}\label{prop:intervalpart}
Given $K\ge 2$, we assume $\frac{n_1}{m_1}, \frac{n_2}{m_2}, \frac{n_3}{m_3}, \frac{n_4}{m_4}\in \mathfrak{S}_K$ and $\frac{n_1}{m_1}< \frac{n_2}{m_2}<\frac{n_3}{m_3}<\frac{n_4}{m_4}$. If $ \frac{n_2}{m_2}$ and $\frac{n_3}{m_3}$ form one Farey pair, then $\frac{n_1+n_3}{m_1+m_3} \le \frac{n_2}{m_2}$ and $\frac{n_3}{m_3} \le \frac{n_2+n_4}{m_2+m_4}$.
~\hfill$\blacksquare$
\end{property}

\subsection{The Minimum Euclidean Distance of the Received Signal Constellation Points}
We are now ready to solve the problem in~\eqref{eqn:mindistD1} to find the constellation point pairs $(m,n)$ that have the minimum Euclidean distance.
To that end, we first have the following preliminary propositions.

\begin{proposition}\label{prop:mindistant}
Let $\mathbb F_{K}^2 =\{(m,n): \frac{n}{m}\in \mathfrak{S}_{K} \}$, where $\mathfrak{S}_{K}$ is the extended Farey number sequence of order $K$, and then we have
\begin{align*}
\min_{\quad (m,n)\in \mathbb Z_{K}^2 \setminus\{(0,0)\} }~d_1(m,n)=\min_{\quad (m,n)\in \mathbb F_{K}^2 }~d_1(m,n).
\end{align*}
\hfill$\blacksquare$
\end{proposition}
The proof is given in Appendix\ref{appendix:propfareyint}.

\begin{proposition}\label{prop:basicfareyextended}
Consider the Farey interval $\big(\frac{n_1}{m_1}, \frac{n_2}{m_2}\big) \in \mathcal{S}_K$, with $K\ge 1$, and $\frac{n_1}{m_1}< \frac{n_2}{m_2}$. Then, for $\frac{|h_{21}| w_2}{|h_{11}|w_1} \in (\frac{n_1}{m_1}, \frac{n_2}{m_2})$ and $d_1(m,n) =\big||h_{11}|w_1 n-|h_{21}| w_2 m\big|$, we have	
	\begin{enumerate}
		\item If $\frac{|h_{21}| w_2}{|h_{11}|w_1}= \frac{n_1 +n_2}{m_1 + m_2}$, then $d_1(m_1, n_1)=d_1(m_2, n_2)$;
		\item If $\frac{|h_{21}| w_2}{|h_{11}|w_1} \in \big(\frac{n_1}{m_1}, \frac{n_1 +n_2}{m_1 + m_2}\big)$, then $d_1(m_1, n_1) <d_1(m_2, n_2)$;
		\item If  $\frac{|h_{21}| w_2}{|h_{11}|w_1} \in \big(\frac{n_1 +n_2}{m_1 + m_2}, \frac{n_2}{m_2}\big)$, then $d_1(m_1, n_1)>d_1(m_2, n_2)$. ~\hfill$\blacksquare$
	\end{enumerate}
\end{proposition}
The proof can be found in Appendix\ref{appendix:fareydiv}.
\begin{proposition}\label{prop:worstcase}
Consider $\frac{n_1}{m_1},\frac{n_2}{m_2},\frac{n_3}{m_3},\frac{n_4}{m_4} \in \mathfrak{S}_{K}$, with $K\ge 2$, such that $\frac{n_1}{m_1}<\frac{n_2}{m_2}<\frac{n_3}{m_3}<\frac{n_4}{m_4} $ where $\frac{n_2}{m_2}, \frac{n_3}{m_3}$ form one Farey pair,
\begin{enumerate}
	\item If $\frac{|h_{21}| w_2}{|h_{11}|w_1} \in (\frac{n_2}{m_2},\frac{n_2+n_3}{m_2+m_3})$, then
	\begin{multline*}
		\min_{(m,n) \in \mathbb F_{M-1}^2}~d_1(m,n)=d_1(m_2, n_2)\\
		=|h_{21}|w_2 m_2- |h_{11}| w_1 n_2.
	\end{multline*}
	\item If  $\frac{|h_{21}| w_2}{|h_{11}|w_1} \in (\frac{n_2+n_3}{m_2+m_3},\frac{n_3}{m_3})$, then
	\begin{multline*}
	\min_{(m,n) \in \mathbb F_{M-1}^2}~d_1(m,n)=d_1(m_3, n_3)\\
		 =|h_{11}| w_1 n_3 - |h_{21}|w_2 m_3.
	\end{multline*}
~\hfill$\blacksquare$	
\end{enumerate}
\end{proposition}
The proof is provided in Appendix\ref{appendix:worstcase}.
\begin{proposition}\label{prop:monotonic}	
For the Farey interval set $\mathcal{S}_{M-1}=\big\{\big(\frac{b_k}{a_k}, \frac{b_{k+1}}{a_{k+1}}\big)\big\}_{k=1}^{|\mathcal{S}_{M-1}|}$, if $\frac{|h_{21}|}{|h_{22}|} \ge b_k+b_{k+1}$, then $\frac{a_{k+1}}{b_{k+1}}+\frac{|h_{22}| }{b_{k+1}|h_{21}|} \le \frac{a_k+a_{k+1}}{b_k+b_{k+1}}\le \frac{a_{k}}{b_{k}}-\frac{|h_{22}| }{b_{k}|h_{21}|} $.~\hfill$\blacksquare$	
\end{proposition}
The proof is given in Appendix\ref{appendix:monotonic}.
\subsection{Optimal Solution to Problem~\ref{pbm:maxminopt} for  $\frac{|h_{21}| w_2}{|h_{11}|w_1} $ in Certain Farey Interval}
In this section, we solve Problem~\ref{pbm:maxminopt} by restricting~$\frac{|h_{21}|w_2 }{|h_{11}| w_1}$ into a certain Farey interval where a closed-form solution is attainable. We consider the Farey interval set $\mathcal{S}_{M-1}$ given by $\mathcal{S}_{M-1}=\big\{\big(\frac{b_k}{a_k},\frac{b_{k+1}}{a_{k+1}}\big)\big\}_{k=1}^{|\mathcal{S}_{M-1}|}$ where $|\mathcal{S}_{M-1}|=2\sum_{m=1}^{M-1} \varphi(m)$. Now we consider the case $\frac{|h_{21}|w_2 }{|h_{11}| w_1} \in \big(\frac{b_k}{a_k}, \frac{b_{k+1}}{a_{k+1}}\big)$,     $k=1,2,\ldots,|\mathcal{S}_{M-1}|$ and we aim to find the optimal
$(w_1^*(k), w_2^*(k))$ such that
\begin{subequations}\label{eqn:maxminbyinterval}
\begin{align}
&g\Big(\frac{b_k}{a_k}, \frac{b_{k+1}}{a_{k+1}}\Big) = \max_{(w_1, w_2)}~\min \Big\{\min_{(m,n) \in \mathbb F_{M-1}^2} d_1(m,n),\nonumber\\
&\qquad\qquad\qquad\qquad\qquad \qquad\qquad\min_{m \in \mathbb Z \setminus \{0\}} d_2(m)\Big\} \\
&{\rm s.t.~}\frac{b_k}{a_k}< \frac{|h_{21}|w_2 }{|h_{11}| w_1}\le \frac{b_{k+1}}{a_{k+1}},0< w_1 \le \sqrt{\frac{3 p_1}{M^2-1}} \nonumber\\
&\qquad\qquad\qquad\qquad\qquad{~\rm~and~} 0 < w_2 \le \sqrt{\frac{3 p_2}{M^2-1}}.
\end{align}
\end{subequations}

By applying the propositions in last subsections, we manage to attain the following lemma in terms of the optimal solution to problem~\eqref{eqn:maxminbyinterval}.
\begin{lemma}\label{thm:subinterval}
The optimal solution to~\eqref{eqn:maxminbyinterval} is given as follows.
	\begin{enumerate}
		\item  If $\frac{|h_{21}|}{|h_{22}|} \le b_k +b_{k+1}$, the following statements are true:
		\begin{enumerate}
			\item If $\frac{|h_{11}|}{|h_{21}|}\ge \frac{\sqrt{p_2}(a_k+a_{k+1})}{\sqrt{p_1}(b_k+b_{k+1})}$, then $g\big(\frac{b_k}{a_k}, \frac{b_{k+1}}{a_{k+1}}\big)=\frac{|h_{21}|}{b_k+b_{k+1}} \sqrt{\frac{3 p_2}{M^2-1}}$ and
			$(w_1^*(k), w_2^*(k))=\Big(\frac{(a_k + a_{k+1})|h_{21}|}{(b_k+b_{k+1}) |h_{11}|}\sqrt{\frac{3 p_2}{M^2-1}}, \sqrt{\frac{3 p_2}{M^2-1}}\Big)$.
			\item If $\frac{|h_{11}|}{|h_{21}|}<\frac{\sqrt{p_2}(a_k+a_{k+1})}{\sqrt{p_1}(b_k+b_{k+1})}$,  then $g\big(\frac{b_k}{a_k}, \frac{b_{k+1}}{a_{k+1}}\big)=\frac{|h_{11}|}{a_k+ a_{k+1}}\sqrt{\frac{3 p_1}{M^2-1}}$ and $(w_1^*(k), w_2^*(k))=\Big(\sqrt{\frac{3 p_1}{M^2-1}}, \frac{(b_k +b_{k+1})|h_{11}|}{(a_k+a_{k+1})|h_{21}|}\sqrt{\frac{3 p_1}{M^2-1}}\Big)$ .
		\end{enumerate}
		\item  If $\frac{|h_{21}|}{|h_{22}|} > b_k +b_{k+1}$, we have the following results:
		\begin{enumerate}
			\item If $\frac{|h_{11}|}{|h_{21}|}\ge  \frac{\sqrt{p_2}}{\sqrt{p_1}} \big(\frac{a_{k+1}}{b_{k+1}}
			+\frac{|h_{22}| }{b_{k+1}|h_{21}|}\big)$,
			then  $g\big(\frac{b_k}{a_k}, \frac{b_{k+1}}{a_{k+1}}\big)=|h_{22}|\sqrt{\frac{3 p_2}{M^2-1}}$ and  $(w_1^*(k), w_2^*(k))=\Big(\frac{a_{k+1} |h_{21}|+|h_{22}| }{ b_{k+1} |h_{11}|} \sqrt{\frac{3 p_2}{M^2-1}}, \sqrt{\frac{3 p_2}{M^2-1}}\Big)$.
			\item  If $\frac{|h_{11}|}{|h_{21}|}<\frac{\sqrt{p_2 }}{\sqrt{p_1}} \big(\frac{a_{k+1}}{b_{k+1}}
			+\frac{|h_{22}| }{b_{k+1}|h_{21}|}\big)$,
			then $g\big(\frac{b_k}{a_k}, \frac{b_{k+1}}{a_{k+1}}\big)=\frac{b_{k+1}|h_{11}||h_{22}|}{a_{k+1}|h_{21}|+|h_{22}|}  \sqrt{\frac{3 p_1}{M^2-1}}$ and $(w_1^*(k), w_2^*(k))=\Big(\sqrt{\frac{3 p_1}{M^2-1}}, \frac{b_{k+1} |h_{11}|}{a_{k+1} |h_{21}|+|h_{22}|}\sqrt{\frac{3 p_1}{M^2-1}}\Big)$ .
		\end{enumerate}
	\end{enumerate}\hfill$\blacksquare$
\end{lemma}
The proof of Lemma~1 can be found in~Appendix\ref{appendix:theorem1}.
\begin{remark} We have the following insights from the above lemma,
\begin{enumerate}
\item It can be observed from Lemma~\ref{thm:subinterval} that at least one transmitter should employ the maximum allowable power, since otherwise we could scale up both transmitted power without violating the power constraint such that the minimum Euclidean distance is enlarged.
\item
We can see from Lemma~\ref{thm:subinterval} that the optimal value of the objective function together with $(w_1^*(k), w_2^*(k))$  substantially depend on the relative strength of the channel coefficients. Inspired by~\cite{Chong07}, we divide the considered real Gaussian ZC into three scenarios:
	\begin{enumerate}
	\item  Gaussian ZC with \emph{weak cross link}, i.e., $\frac{|h_{21}|}{|h_{22}|} \in (0,1]$;
	    \item  Gaussian ZC with~\emph{strong cross link}, i.e., $\frac{|h_{21}|}{|h_{22}|} \in (1,2M)$;
	\item  Gaussian ZC with \emph{very strong cross link}, i.e., $\frac{|h_{21}|}{|h_{22}|} \in [2M, \infty)$.
	\end{enumerate}
Then, a compact closed-form expression for Problem~\ref{pbm:maxminopt} can be established for three complimentary scenarios, which constitutes main contents of the subsequent subsection.~\hfill\QED
\end{enumerate}
\end{remark}

\subsection{The Optimal NOMA Design with PAM Constellation for the Gaussian ZC}
Now we are ready to give the closed-form optimal solution of $(w_1^*, w_2^*)$  to Problem~\ref{pbm:maxminopt} which maximizes the minimum Euclidean distance over all the Farey intervals for the aforementioned three scenarios.

\subsubsection{{\bf Scenario~1:} ZC with Weak Cross Link}
For this case, we have $\frac{|h_{21}|}{|h_{22}|} \in (0,1]$. Consider the Farey interval set $\mathcal{S}_{M-1}=\big\{\big(\frac{b_k}{a_k}, \frac{b_{k+1}}{a_{k+1}}\big)\big\}_{k=1}^{|\mathcal{S}_{M-1}|}$. By Property~\ref{prop:basicFareyprop}, we have $\frac{a_{k+1}+a_{k+2}}{b_{k+1}+b_{k+2}}<\frac{a_{k+1}}{b_{k+1}} <\frac{a_k+a_{k+1}}{b_k+b_{k+1}}$, and therefore the positive real axis can be divided into the $|\mathcal{S}_{M-1}|+1$ intervals in an increasing order:
\begin{align*}
	&\Bigg\{\Big(0, \frac{a_{|\mathcal{S}_{M-1}|}+a_{|\mathcal{S}_{M-1}|+1}  }{b_{|\mathcal{S}_{M-1}|}+b_{|\mathcal{S}_{M-1}|+1} }\Big),\Big(\frac{a_{|\mathcal{S}_{M-1}|}+a_{|\mathcal{S}_{M-1}|+1}  }{b_{|\mathcal{S}_{M-1}|}+b_{|\mathcal{S}_{M-1}|+1} },\\
	&\qquad \qquad\qquad\frac{a_{|\mathcal{S}_{M-1}|-1} +a_{|\mathcal{S}_{M-1}|} }{b_{|\mathcal{S}_{M-1}|-1} +b_{|\mathcal{S}_{M-1}|}}\Big), \cdots,\Big(\frac{a_{1} +a_{2} }{b_{1} +b_{2}},\infty \Big)\Bigg\}.
\end{align*}
In particular, we have $\frac{a_{1} +a_{2} }{b_{1} +b_{2}}=M$ and $\frac{a_{|\mathcal{S}_{M-1}|}+a_{|\mathcal{S}_{M-1}|+1}  }{b_{|\mathcal{S}_{M-1}|}+b_{|\mathcal{S}_{M-1}|+1} }=\frac{1}{M}$.

\begin{theorem}[Gaussian ZC with weak cross link]\label{thm:weakint}
Suppose that $\frac{|h_{21}|}{|h_{22}|} \in (0,1]$.
Then, the optimal power scaling factors to Problem~\ref{pbm:maxminopt} are explicitly determined as follows:
	\begin{enumerate}
		\item If $\frac{|h_{11}|}{|h_{21}|} \le \frac{\sqrt{p_2}}{M\sqrt{p_1}}$, then, we have $(w_1^*, w_2^*)=\Big(\sqrt{\frac{3 p_1}{M^2-1}}, \frac{M |h_{11}|}{|h_{21}|}\sqrt{\frac{3 p_1}{M^2-1}}\Big)$;
		
		\item If $\frac{|h_{11}|}{|h_{21}|} \ge \frac{M \sqrt{ p_2} }{\sqrt{p_1}}$, then, we have  $(w_1^*, w_2^*)=\Big(\frac{M |h_{21}|}{|h_{11}|}\sqrt{\frac{3 p_2}{M^2-1}},\sqrt{\frac{3 p_2}{M^2-1}}\Big)$;
		
		\item Let $\frac{|h_{11}|}{|h_{21}| } \in \Big(\frac{\sqrt{p_2 }(a_{\ell_1+1} +a_{\ell_1+2})  }{\sqrt{p_1}(b_{\ell_1+1} +b_{\ell_1+2})}, \frac{ \sqrt{p_2}(a_{\ell_1} +a_{\ell_1+1}) }{\sqrt{p_1}(b_{\ell_1} +b_{\ell_1+1})}\Big)$ for some $\ell_1=1, \cdots,   |\mathcal{S}_{M-1}|-1$. If we let $\tilde{\ell}_a=\arg \min_k \{(a_1+a_2), \cdots, (a_{\ell_1}+ a_{\ell_1+1}) \}$ and
		$\tilde{\ell}_b=\arg \min_k \{(b_{\ell_1+1}+b_{\ell_1+2}), \cdots, (b_{|\mathcal{S}_{M-1}|}+b_{|\mathcal{S}_{M-1}|+1})\}$, then we have
		\begin{align*}
			&(w_1^*, w_2^*)=\begin{cases}
				\Big(\sqrt{\frac{3 p_1}{M^2-1}},\frac{(b_{\tilde{\ell}_a} +b_{\tilde{\ell}_a+1})|h_{11}|}{(a_{\tilde{\ell}_a}+a_{\tilde{\ell}_a+1})|h_{21}|}\sqrt{\frac{3 p_1}{M^2-1}}\Big),\\ \qquad\qquad {~\rm~if~} \frac{|h_{11}|}{|h_{21}|} \ge \frac{\sqrt{ p_2}(a_{\tilde{\ell}_a}+ a_{\tilde{\ell}_a+1}) }{\sqrt{ p_1}(b_{\tilde{\ell}_b}+b_{\tilde{\ell}_b+1})};\\
				\Big(\frac{(a_{\tilde{\ell}_b} + a_{\tilde{\ell}_b+1})|h_{21}|}{(b_{\tilde{\ell}_b}+b_{\tilde{\ell}_b+1}) |h_{11}|}\sqrt{\frac{3 p_2}{M^2-1}}, \sqrt{\frac{3 p_2}{M^2-1}}\Big),\\
				\qquad\qquad {~\rm~if~} \frac{|h_{11}|}{|h_{21}|} < \frac{\sqrt{ p_2}(a_{\tilde{\ell}_a}+ a_{\tilde{\ell}_a+1}) }{\sqrt{ p_1}(b_{\tilde{\ell}_b}+b_{\tilde{\ell}_b+1})}.
			\end{cases}
		\end{align*}
	\end{enumerate}
	\hfill\QED
\end{theorem}	
The proof of Theorem~\ref{thm:weakint} is provided in~Appendix\ref{appendix:theoremweakint}.
\subsubsection{{\bf Scenario~2:} ZC with Strong Cross Link}
In this case,  $\frac{|h_{21}|}{|h_{22}|} \in (1, 2M)$. We suppose that $L-1 <\frac{|h_{21}|}{|h_{22}|} \le L$.
Then, the optimal solution to Problem~1 can be obtained by considering the following two cases: the Farey interval set $\mathcal{U}_{M-1}^L=\{\big(\frac{b_k'}{a_k'}, \frac{b_{k+1}'}{a_{k+1}'}\big)\in \mathcal{S}_{M-1}: b_k' +b_{k+1}' \ge L\}$ and $\mathcal{V}_{M-1}^L=\{\big(\frac{b_k''}{a_k''}, \frac{b_{k+1}''}{a_{k+1}''}\big)\in \mathcal{S}_{M-1}: b_k'' +b_{k+1}'' < L\}$. The whole discussions on them can be summarized as the following theorem.
\begin{theorem}[ZC with Strong Cross Link] The optimal solution to Problem~\ref{pbm:maxminopt} in this case can be obtained by finding the maximum value of the objective functions in the following two cases which are explicitly attained as follows:
\begin{enumerate}
\item
Let $\big(\frac{b_k'}{a_k'}, \frac{b_{k+1}'}{a_{k+1}'}\big)\in\mathcal{U}_{M-1}^L$. Then, the following three statements are true.	\begin{enumerate}
		\item If $\frac{|h_{11}|}{|h_{21}|} \le \frac{\sqrt{p_2}(a_{|\mathcal{U}_{M-1}^L|}' +a_{|\mathcal{U}_{M-1}^L|+1}' )}{\sqrt{p_1}(b_{|\mathcal{U}_{M-1}^L|}' +b_{|\mathcal{U}_{M-1}^L|+1}')}$, then we have $(w_{u_1}^*, w_{u_2}^*)=\Big(\sqrt{\frac{3 p_1}{M^2-1}}, \frac{(b_{\tilde{\ell}_c}'+b_{\tilde{\ell}_c+1}') |h_{11}|}{(a_{\tilde{\ell}_c}'+a_{\tilde{\ell}_c+1}')|h_{21}|}\sqrt{\frac{3 p_1}{M^2-1}}\Big)$, where $\tilde{\ell}_c=\arg\min_{k}\{(a_{1}'+a_{2}'), \ldots, (a_{|\mathcal{U}_{M-1}^L|}'+a_{|\mathcal{U}_{M-1}^L|+1}') \}$.
		
		\item If $\frac{|h_{11}|}{|h_{21}|} \ge \frac{ M\sqrt{p_2} }{\sqrt{p_1}}$, then we have  $(w_{u_1}^*, w_{u_2}^*)=\Big(\frac{M |h_{21}|}{|h_{11}|}\sqrt{\frac{3 p_2}{M^2-1}},\sqrt{\frac{3 p_2}{M^2-1}}\Big)$.
		\item Suppose that $\frac{|h_{11}|}{|h_{21}| } \in \Big(\frac{\sqrt{p_2}(a_{\ell_2+1}' +a_{\ell_2+2}')  }{\sqrt{p_1 }(b_{\ell_2+1}' +b_{\ell_2+2}')}, \frac{ \sqrt{p_2}(a_{\ell_2}' +a_{\ell_2+1}') }{\sqrt{p_1}(b_{\ell_2}' +b_{\ell_2+1}')}\Big)$ for some $\ell_2'=1, \cdots,  |\mathcal{U}_{M-1}^L|$. If we let $\tilde{\ell}_d=\arg \min_k \{(a_1+a_2), \ldots, (a_{\ell_2}+ a_{\ell_2+1}) \}$ and
		$\tilde{\ell}_e=\arg \min_k \{(b_{\ell_2+1}+b_{\ell_2+2}),\ldots, (b_{|\mathcal{U}_{M-1}^L|}+b_{|\mathcal{U}_{M-1}^L|+1})\}$, then we have
		\begin{align*}
			(w_{u_1}^*, w_{u_2}^*)\!\!=\!\!\begin{cases}
				\!\!\big(\sqrt{\frac{3 p_1}{M^2-1}},\frac{(b_{\tilde{\ell}_d}' +b_{\tilde{\ell}_d+1}')|h_{11}|}{(a_{\tilde{\ell}_d}'+a_{\tilde{\ell}_d+1}')|h_{21}|}\sqrt{\frac{3 p_1}{M^2-1}}\big),\\
				{\rm~if~}\frac{|h_{11}|}{|h_{21}|} \ge \frac{\sqrt{ p_2}(a_{\tilde{\ell}_d}'+ a_{\tilde{\ell}_d+1}') }{\sqrt{ p_1}(b_{\tilde{\ell}_e}'+b_{\tilde{\ell}_e+1}')};\\
				\!\!\big(\frac{(a_{\tilde{\ell}_e}' + a_{\tilde{\ell}_e+1}')|h_{21}|}{(b_{\tilde{\ell}_e}'+b_{\tilde{\ell}_e+1}') |h_{11}|}\sqrt{\frac{3 p_2}{M^2-1}}, \sqrt{\frac{3 p_2}{M^2-1}}\big)\\
				{\rm~if~}\frac{|h_{11}|}{|h_{21}|} < \frac{\sqrt{ p_2}(a_{\tilde{\ell}_d}'+ a_{\tilde{\ell}_d+1}') }{\sqrt{ p_1}(b_{\tilde{\ell}_e}'+b_{\tilde{\ell}_e+1}')}.
			\end{cases}
		\end{align*}
	\end{enumerate}
	
	\item  Let $\big(\frac{b_{k-1}''}{a_{k-1}''}, \frac{b_{k}''}{a_{k}''}\big) \in {\mathcal{V}}_{M-1}^L$. Then, the following two statements are true.	
	\begin{enumerate}
		\item If $\frac{|h_{11}|}{|h_{22}|} \le \frac{ \sqrt{p_2}}{\sqrt{p_1}} $, then we have $(w_{v_1}^*, w_{v_2}^*)=( \sqrt{\frac{3 p_1}{M^2-1}}, \frac{b_{|\mathcal{V}_{M-1}|+1}'' |h_{11}||h_{22}|}{a_{|\mathcal{V}_{M-1}|+1}'' |h_{21}|+|h_{22}|}  \sqrt{\frac{3 p_1}{M^2-1}})$.
		\item If 
		$\frac{|h_{11}|\sqrt{p_1}}{|h_{21}| \sqrt{p_2}} \in \big(\frac{a_{\ell_3+1}''}{b_{\ell_3+1}''}+\frac{|h_{22}| }{b_{\ell_3+1}'' |h_{21}|}, \frac{a_{\ell_3}''}{b_{\ell_3}''}+\frac{|h_{22}| }{b_{\ell_3}'' |h_{21}|}\big)$ for some $\ell_3=1,\cdots, |\mathcal{V}_{M-1}|$, then we have
		\begin{align*}
			(w_{v_1}^*,w_{v_2}^*)=\begin{cases}
				\big(\sqrt{\frac{3 p_1}{M^2-1}}, \frac{b_{\ell_3}''|h_{11}|}{a_{\ell_3}''|h_{21}|+|h_{22}|}\sqrt{\frac{3 p_1}{M^2-1}}\big),\\
				{\rm~if~}\frac{|h_{11}|}{|h_{21}|}\ge \frac{\sqrt{p_2 }}{\sqrt{p_1 }}\big(\frac{a_{{\ell_3}}''}{b_{\ell_3}''}+\frac{|h_{22}| }{b_{\ell_3}''|h_{21}|}\big);\\
				\big(\frac{a_{\ell_3+1}''|h_{21}|+|h_{22}|}{b_{\ell_3+1}'' |h_{11}| }  \sqrt{\frac{3 p_2}{M_2^2-1}}, \sqrt{\frac{3 p_2}{M_2^2-1}}\big)\\{\rm~if~}\frac{|h_{11}|}{|h_{21}|}< \frac{\sqrt{p_2}}{\sqrt{p_1}}\big(\frac{a_{{\ell_3}''}}{b_{\ell_3}''}+\frac{|h_{22}| }{b_{\ell_3}''|h_{21}|}\big).
			\end{cases}
		\end{align*}	
	\end{enumerate}
\end{enumerate}
~\hfill\QED
\end{theorem}
The proof of Theorem~2  is  very similar to that of Theorem~1 and the following Theorem~3 and thus, is omitted due to space limitation.
\subsubsection{{\bf Scenario~3:} ZC with Very Strong Cross Link}
In this case, $\frac{|h_{21}|}{|h_{22}|} \in [2M, \infty)$. Likewise, we consider the Farey interval set $\mathcal{S}_{M-1}=\big\{\big(\frac{b_k}{a_k}, \frac{b_{k+1}}{a_{k+1}}\big)\big\}_{k=1}^{|\mathcal{S}_{M-1}|}$. Note that $\frac{|h_{21}|}{|h_{22}|} \ge 2M> b_{k}+b_{k+1}$ for $k=1,\ldots,|\mathcal{S}_{M-1}|$. Then, by Property~\ref{prop:basicFareyprop} and Proposition~\ref{prop:monotonic}, we have
$\frac{a_{k+1}}{b_{k+1}}+\frac{|h_{22}| }{b_{k+1}|h_{21}|} \le \frac{a_k+a_{k+1}}{b_k+b_{k+1}}< \frac{a_{k}}{b_{k}} <\frac{a_{k}}{b_{k}}+\frac{|h_{22}| }{b_{k}|h_{21}|} $. As a result, the positive real axis can be divided into the following intervals in increasing order:
\begin{small}
\begin{multline*}
	\Big\{\big(0, \frac{a_{|\mathcal{S}_{M-1}|+1}}{b_{|\mathcal{S}_{M-1}|+1}}+\frac{|h_{22}| }{b_{|\mathcal{S}_{M-1}|+1}|h_{21}|}\big),\\
	\ldots, \big(\frac{a_{2}}{b_{2}}+\frac{|h_{22}| }{b_{2} |h_{21}|}, \frac{a_{1}}{b_{1}}+\frac{|h_{22}| }{b_{1} |h_{21}|}\big)\Big\},
\end{multline*}
\end{small}
where $\frac{a_{1}}{b_{1}}+\frac{|h_{22}| }{b_{1} |h_{21}|}=\infty$.
\begin{theorem}[Gaussian ZC with very strong cross link]\label{thm:verystrongint}
	Let $\frac{|h_{21}|}{|h_{22}|} \in [2M, \infty)$. Then, the optimal solution to Problem~\ref{pbm:maxminopt} is given below:
	\begin{enumerate}
		\item If $\frac{|h_{11}|}{|h_{22}|} \le \frac{ \sqrt{p_2}}{\sqrt{p_1}}$, then we have $(w_1^*, w_2^*)=(\sqrt{\frac{3 p_1}{M^2-1}},  \frac{|h_{11}|}{|h_{22}|}\sqrt{\frac{3 p_1}{M^2-1}})$.
		\item If $\frac{|h_{11}|\sqrt{p_1}}{|h_{21}| \sqrt{p_2}} \in \big(\frac{a_{\ell_4+1}}{b_{\ell_4+1}}+\frac{|h_{22}| }{b_{\ell_4+1} |h_{21}|}, \frac{a_{\ell_4}}{b_{\ell_4}}+\frac{|h_{22}| }{b_{\ell_4} |h_{21}|}\big)$ for some $\ell_4=1,\ldots, |\mathcal{S}_{M-1}|$, then we have
		\begin{align*}
			(w_1^*,w_2^*)=\begin{cases}
				\big(\sqrt{\frac{3 p_1}{M^2-1}}, \frac{b_{\ell_4}|h_{11}|}{a_{\ell_4}|h_{21}|+|h_{22}|}\sqrt{\frac{3 p_1}{M^2-1}}\big),\\
				{\rm~if~} \frac{|h_{11}|}{|h_{21}|}\ge \frac{\sqrt{p_2}}{\sqrt{p_1 }}\big(\frac{a_{{\ell_4}}}{b_{\ell_4}}+\frac{|h_{22}| }{b_{\ell_4}|h_{21}|}\big);\\
				\big(\frac{a_{\ell_4+1}|h_{21}|+|h_{22}|}{b_{\ell_4+1} |h_{11}| }  \sqrt{\frac{3 p_2}{M^2-1}}, \sqrt{\frac{3 p_2}{M^2-1}}\big),\\
				{\rm~if~}\frac{|h_{11}|}{|h_{21}|}< \frac{\sqrt{p_2}}{\sqrt{p_1}}\big(\frac{a_{{\ell_4}}}{b_{\ell_4}}+\frac{|h_{22}| }{b_{\ell_4}|h_{21}|}\big).
			\end{cases}
		\end{align*}
	\end{enumerate}\hfill\QED
\end{theorem}
The proof of Theorem~\ref{thm:verystrongint} is provided in Appendix\ref{appendix:theoremverystrongint}.

\section{Simulation Results and Discussions}
%

In this section, computer simulations are carried out to demonstrate the effectiveness of our proposed NOMA design under different channel configurations. More precisely, we compare our proposed NOMA design with CR based NOMA~\cite{Harshan11}, time-division multiple access (TDMA) and frequency-division multiple access (FDMA) approaches. Without loss of generality, we set $P_1=P_2=1$. For simplicity, in the simulations, we assume that the same square-QAM constellation is adopted by both users, i.e., $M=M'$ and  according to~\eqref{eqn:powerallocation}, we have $p_1=p_1'=P_1/2=1/2$ and $p_2=p_2'=P_2/2=1/2$.

\subsection{The Resulting Optimal Sum Constellation at Receiver $D_1$ For Several Deterministic Channels}
\begin{figure*}
	\centering
	\subfigure[]{
		\includegraphics[width=0.46\linewidth]{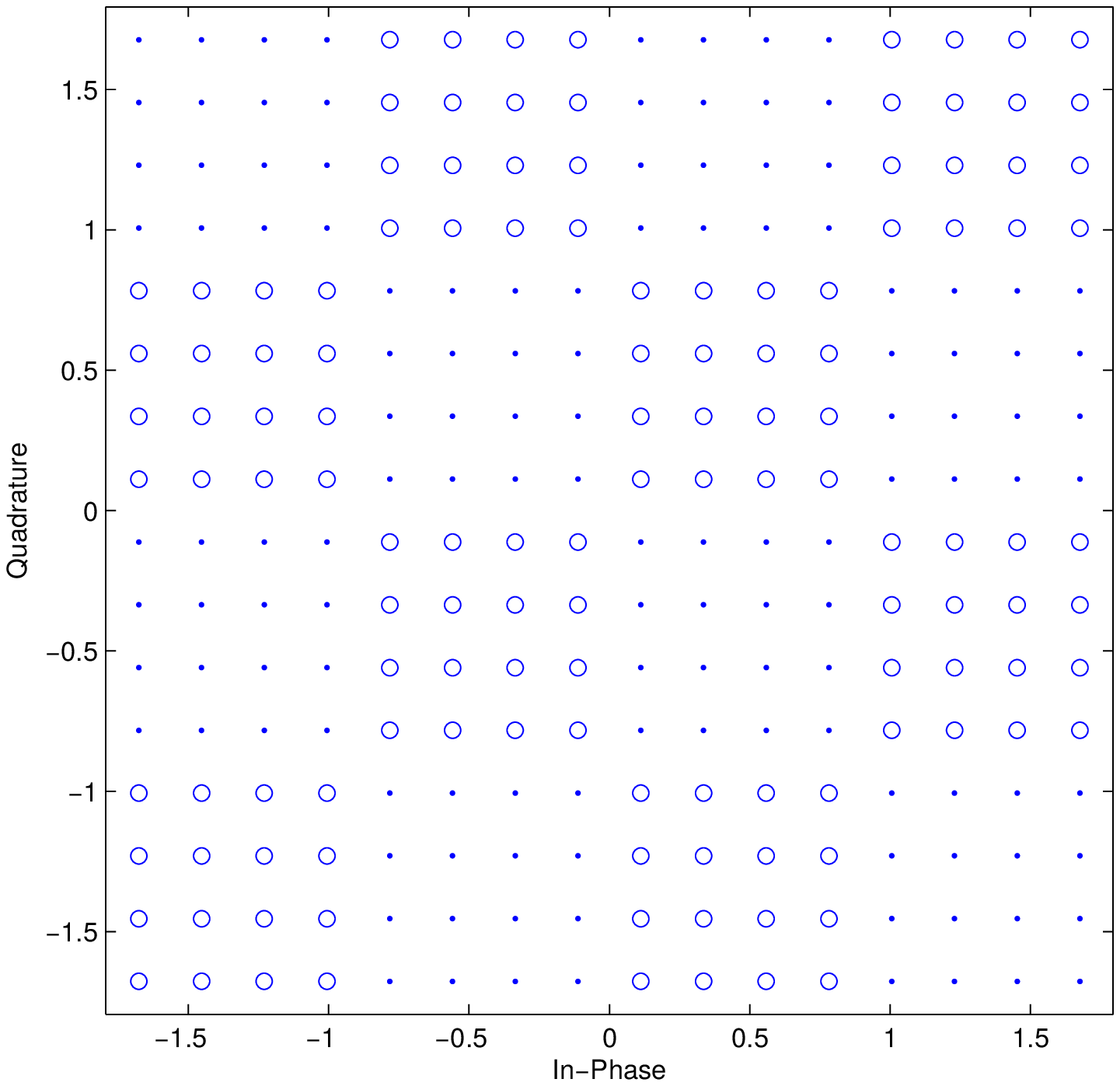}
		\label{fig:weakintsumcons}
	}
	\subfigure[]{
		\includegraphics[width=0.46\linewidth]{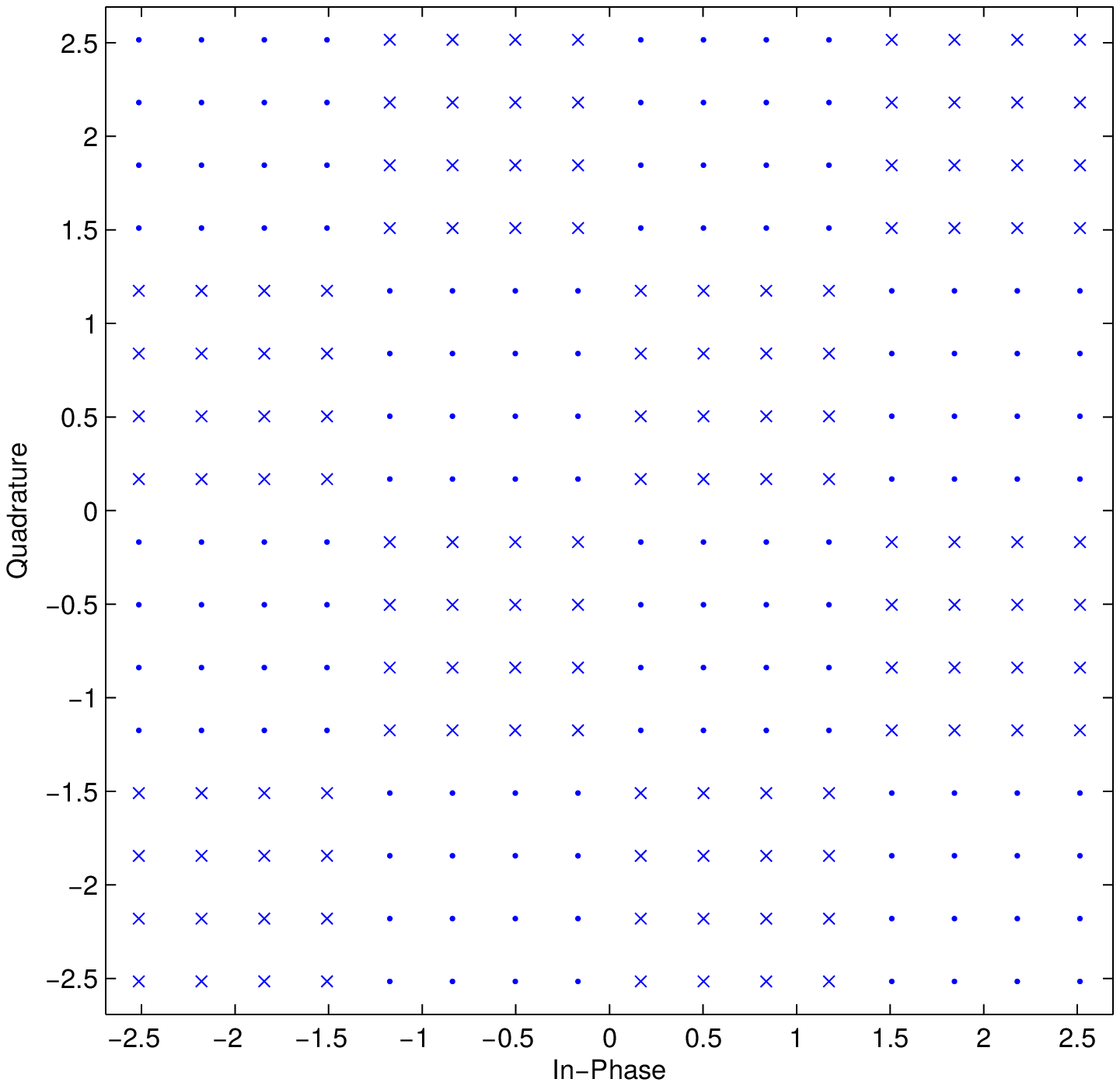}
		\label{fig:strongintsumcons}
	}
	\subfigure[]{
	\includegraphics[width=0.46\linewidth]{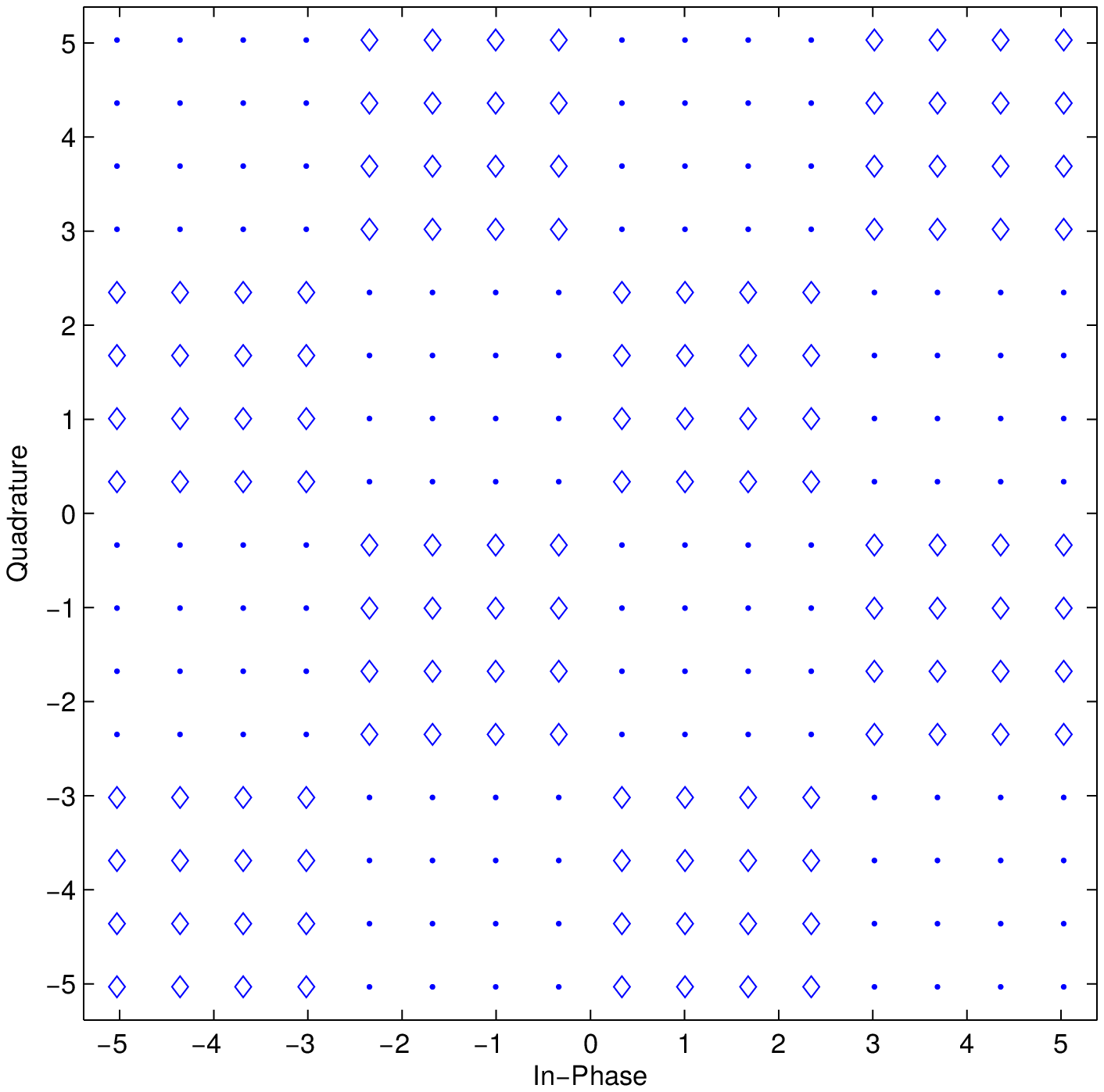}
	\label{fig:verystrintsumconsr1}
}
\subfigure[]{
	\includegraphics[width=0.46\linewidth]{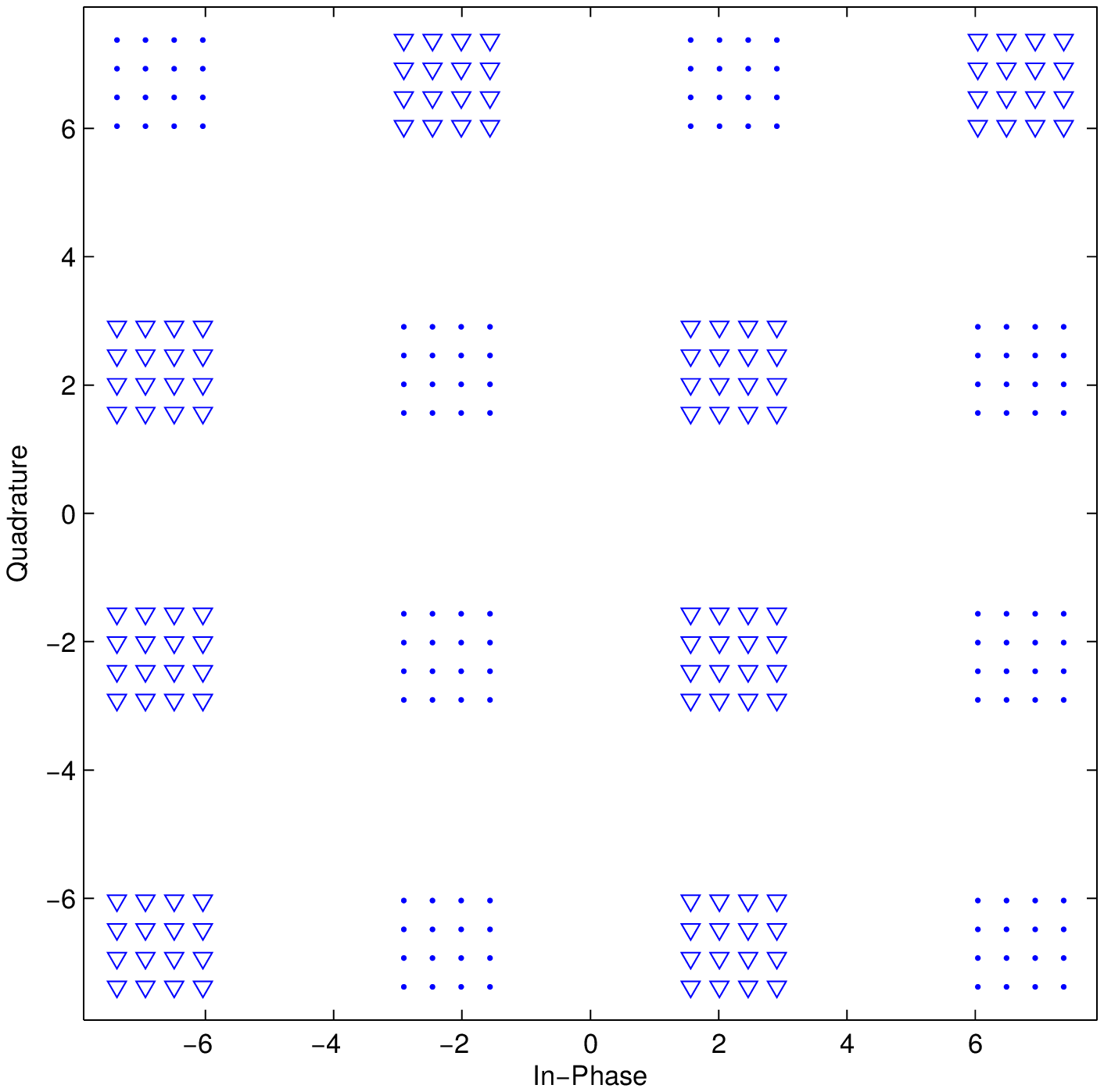}
	\label{fig:verystrintsumconsr2}
}
	\vspace{-10pt}\caption{Optimal resulting constellations at receiver $D_1$ for various cases: (a) Weak cross link with $h_{11}=1$, $h_{21}=1/2$ and $h_{22}=1$.~(b) Strong cross link with $h_{11}=1$, $h_{21}=3/2$ and $h_{22}=1$.
	(c) Strong cross link with $h_{11}=1$, $h_{21}=3$ and $h_{22}=1$ and (d) Very strong cross link with $h_{11}=1$, $h_{21}=5$ and $h_{22}=1/2$.}
\end{figure*}
Here, we consider several deterministic channels corresponding to the three scenarios of \emph{weak cross link}, \emph{strong cross link} and \emph{very strong cross link}. We assume that 16-QAM constellations are used by both users with $M=M'=4$. We discuss the results of these cases one by one as follows:
\subsubsection{Weak Cross Link}
In this case, we assume that $h_{11}=1$, $h_{21}=1/2$ and $h_{22}=1$. Based on the derived expressions for the optimal solution provided in the previous section, we can readily obtain that $(w_1^*, w_2^*)=(0.4472, 0.2236)$.
The corresponding received constellation at $D_1$ is plotted in Fig.\,\ref{fig:weakintsumcons}.
It can be observed that the sum-constellation at $D_1$ is a regular 256-QAM generated by the superposition of two 16-QAM.
Hereafter, we call the signal constellation with smaller minimum Euclidean distance as the satellite constellation. By observing that $w_1^* h_{11}=0.4472$ and $w_2^* h_{21}=0.1118$ (i.e., $w_2^* h_{21}=\frac{1}{4}w_1^* h_{11}$), we can deduce that the constellation used by $S_2$ forms the satellite constellation of the sum constellation at $D_1$.
\subsubsection{Strong Cross Link}
We investigate two channel realizations for this scenario. For the first realization, we let $h_{11}=1$, $h_{21}=3/2$ and $h_{22}=1$. We then have $(w_1^*, w_2^*)=(0.1677, 0.4472)$ and the resulting sum-constellation at $D_1$ is also regular, as illustrated in Fig.\,\ref{fig:strongintsumcons}. Since $w_1^* h_{11}=0.1677$ and $w_2^* h_{21}=0.6708$ (i.e., $w_2^* h_{21}=4 w_1^* h_{11}$) in this case, the constellation used by $S_1$ forms the satellite constellation at $D_1$. For the second realization, we set $h_{11}=1$, $h_{21}=3$ and $h_{22}=1$, leading to $(w_1^*, w_2^*)=(0.3354, 0.4472)$. The resulting constellation plotted in Fig.\,\ref{fig:verystrintsumconsr1} is also uniform as in the previous two scenarios. We have $w_1^* h_{11}=0.3354$ and $w_2^* h_{21}=1.3416$ (i.e., $w_2^* h_{21}=4 w_1^* h_{11}$). Thus, the constellation used by the transmitter $S_1$ forms the satellite constellation at $D_1$.

\subsubsection{Very Strong Cross Link}
In this case, we suppose that $h_{11}=1$, $h_{21}=5$ and $h_{22}=1/2$, generating $(w_1^*, w_2^*)=(0.2236, 0.4472)$. The obtained constellation at $D_1$ is shown in Fig.\,\ref{fig:verystrintsumconsr2}. In this case, we have $w_1^* h_{11}=0.2236$ and $w_2^* h_{21}=2.236$ (i.e., $w_2^* h_{21} =10 w_1^* h_{11}$).
However, it can be observed that $w_2^* h_{22}=0.2236$, i.e., $w_1^* h_{11}=w_2^* h_{22}$.

\subsection{Average Error Performance Comparison in Rayleigh Fading Channels}
We now compare the average BER of the proposed NOMA scheme with that of three existing methods, including TDMA, FDMA, and NOMA with CR~\cite{Harshan11} methods, over Rayleigh fading channels with $h_{11}\sim \mathcal{CN}(0, \delta_{11}^2)$, $h_{21}\sim \mathcal{CN}(0, \delta_{21}^2)$ and $h_{22}\sim \mathcal{CN}(0, \delta_{22}^2)$.
 Recall that we use error performance (i.e., BER) as the design criterion for the NOMA in ZCs with finite-alphabet inputs using \emph{fixed transmission rate} (i.e., fixed constellation size). However, we are unable to compare the error performance of the considered system using finite-alphabet inputs with that of Gaussian inputs. This is because for Gaussian input, it is intractable to evaluate the BER for uncoded system since its input signal is continuous. Moreover, the BER for coded system with Gaussian input is hard to simulate due to the huge storage capacity requirement for the large codebook and the high computational complexity\,\cite[Ch.\,9]{Cover06}.  

For TDMA, we assume that both users transmit alternatively by using half of the total time slots and thus no interference occurs at the destination side. More importantly, the individual instantaneous power constraints on both users $S_1$ and $S_2$ \emph{remain unchanged}. On the other hand, for FDMA, each user uses only half of the available bandwidth. Due to the orthogonality between different frequency band, there is also no interference occurring at the destination side. Note that, in FDMA, the bandwidth occupied by each user is halved and the noise arises at the receiver is assumed to be white Gaussian. Therefore, the variance of the noise is also halved.
In addition, for the  CR based NOMA  as proposed in~\cite{Harshan11}, we let each user transmit at the maximum allowable power by using constellations $\{\exp(\frac{j 2\pi k}{N})\}_{k=0}^{N-1}$ and $\{\exp(\frac{j 2\pi \ell +j\pi}{N})\}_{\ell=0}^{N-1}$ for user $S_1$ and $S_2$, respectively.
\begin{figure}
	\centering
	\subfigure[]{
		\includegraphics[width=0.8\linewidth]{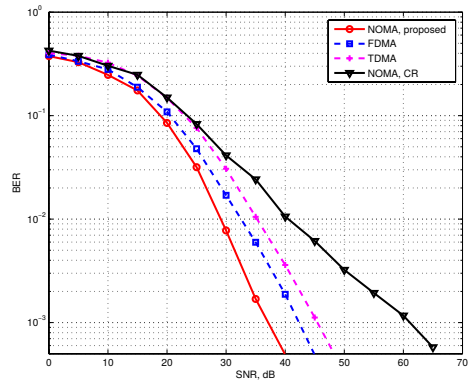}
		\label{fig:comparison16qamcase}
	}
	\subfigure[]{
		\includegraphics[width=0.8\linewidth]{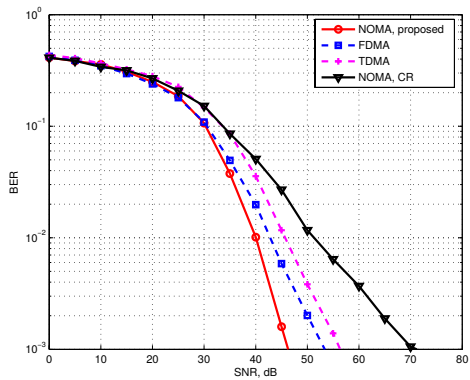}
	\label{fig:comparison64qamcase}
	}
	\vspace{-10pt} {\small} \caption{ Comparison of the proposed NOMA, CR based NOMA, TDMA and FDMA with $(\delta_{11}^2,\delta_{21}^2, \delta_{22}^2)=(1,1,1)$: (a) 16-QAM is used for proposed NOMA while 16-PSK is used for CR based NOMA.~(b) 64-QAM is used for proposed NOMA while 64-PSK is used for CR based NOMA.}\label{fig:stronginterference}
\end{figure}

\begin{figure}
	\centering
	\subfigure[]{
		\includegraphics[width=0.8\linewidth]{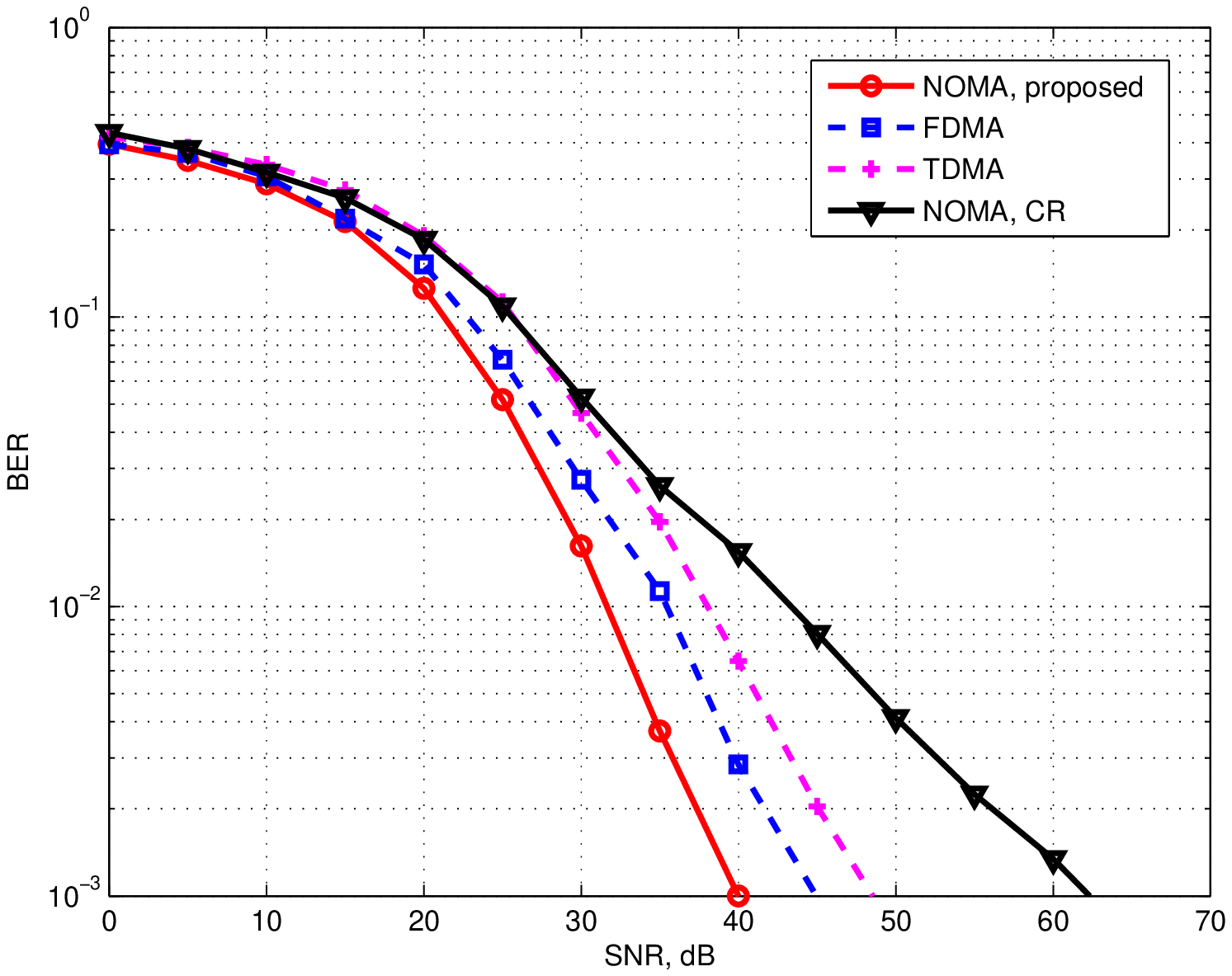}
		\label{fig:NOMA_weak_small}
	}
	\subfigure[]{
		\includegraphics[width=0.8\linewidth]{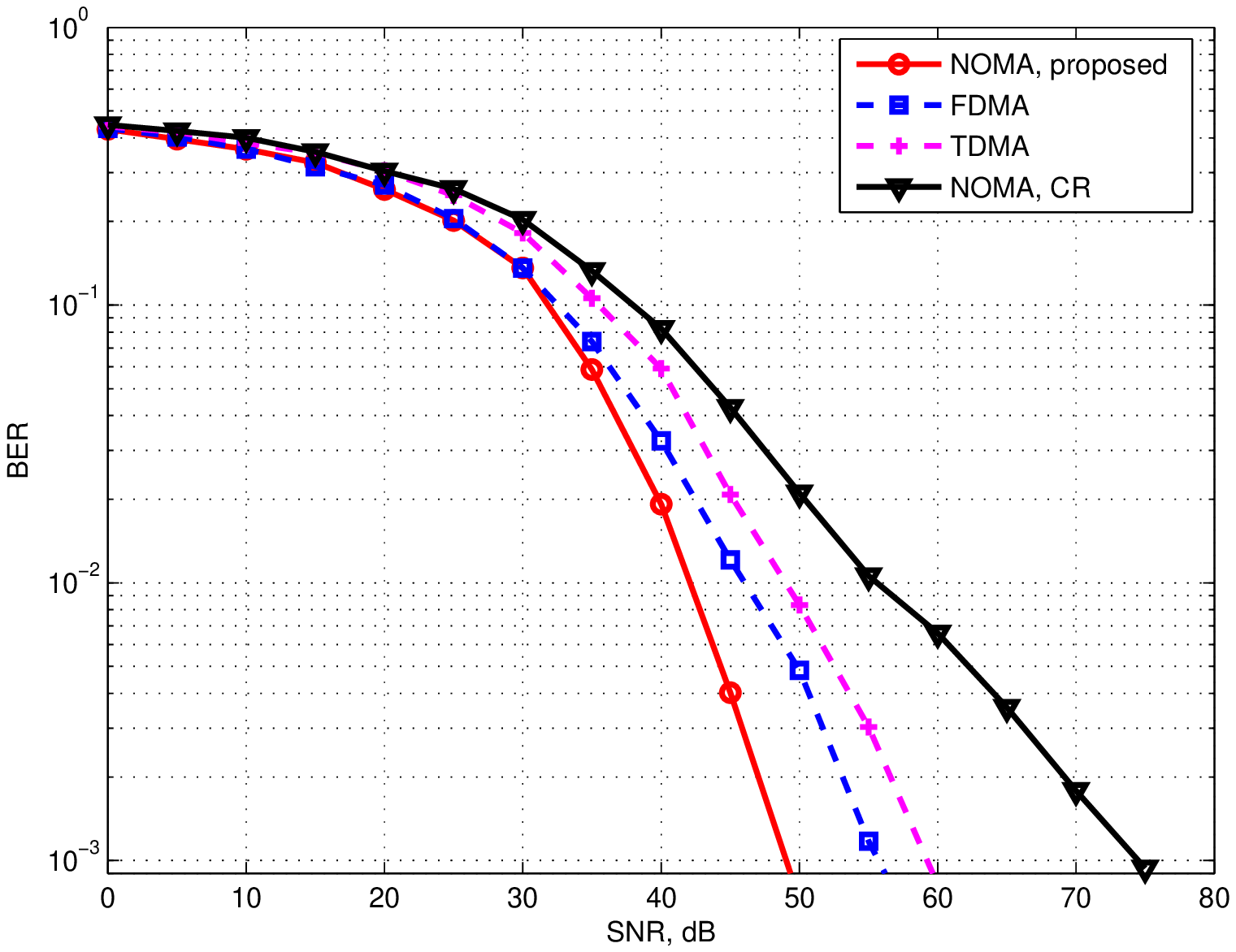}
		\label{fig:NOMA_weak_large}
	}
	\vspace{-10pt}\caption{Comparison of the proposed NOMA, CR based NOMA, TDMA and FDMA with $(\delta_{11}^2,\delta_{21}^2, \delta_{22}^2)=(1,1/4,1)$: (a) 16-QAM is used for proposed NOMA while 16-PSK is used for CR based NOMA.~(b) 64-QAM is used for proposed NOMA while 64-PSK is used for CR based NOMA.}\label{fig:weakinterference}
\end{figure}

\begin{figure}
	\centering
	\subfigure[]{
		\includegraphics[width=0.8\linewidth]{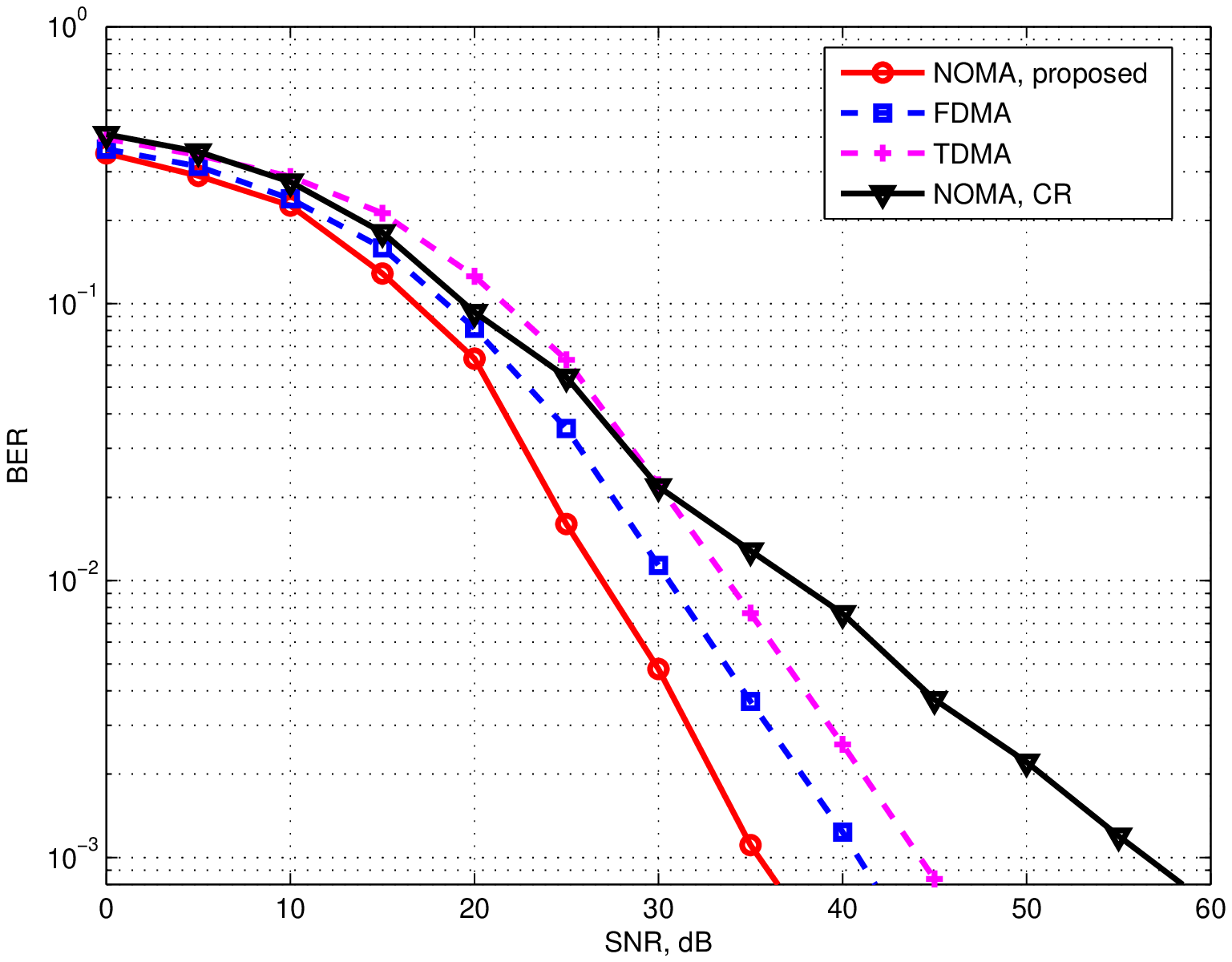}
		\label{fig:NOMA_vstrong_small}
	}
	\subfigure[]{
		\includegraphics[width=0.8\linewidth]{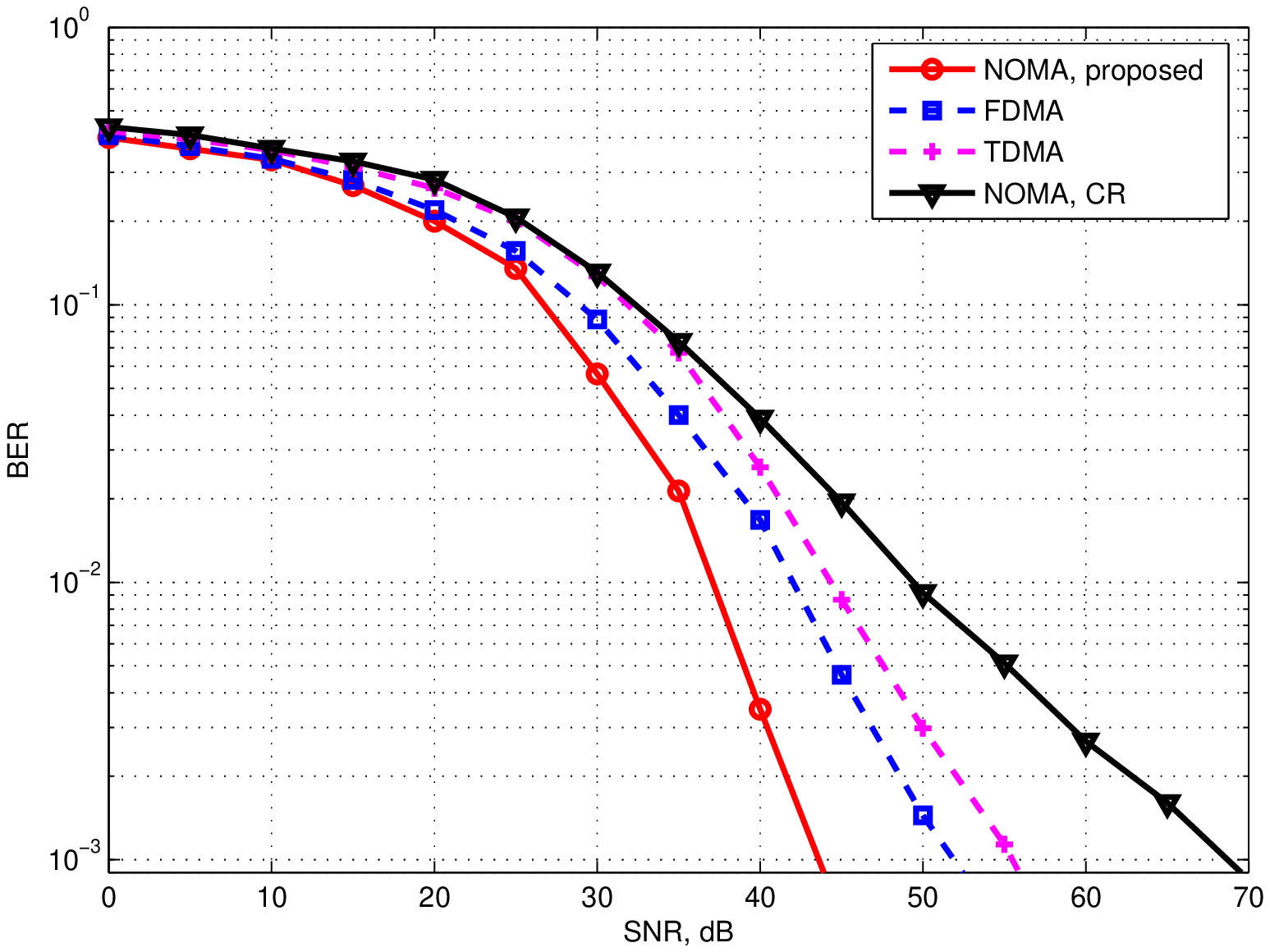}
		\label{fig:NOMA_vstrong_large}
	}
	\vspace{-10pt}\caption{Comparison of the proposed NOMA, CR based NOMA, TDMA and FDMA with $(\delta_{11}^2,\delta_{21}^2, \delta_{22}^2)=(1,4,1)$: (a) 16-QAM is used for proposed NOMA while 16-PSK is used for CR based NOMA.~(b) 64-QAM is used for proposed NOMA while 64-PSK is used for CR based NOMA.}\label{fig:verystronginterference}
\end{figure}

In Fig.\,\ref{fig:stronginterference}, we consider that the variances of all channels are the same, i.e., $(\delta_{11}^2, \delta_{21}^2, \delta_{22}^2)=(1,1,1)$, and the average BER over both receivers of all the methods are plotted against the SNR $\rho=\frac{1}{2\sigma^2}$. In Fig.\,\ref{fig:comparison16qamcase}, 16-QAM is used for the proposed NOMA scheme while 16-PSK is employed by the NOMA with CR. Since only half of the total time slots or total bandwidths are available for each transmitter, to maintain the same data rate for each user in each block compared with NOMA methods, we should increase the constellation size by using 256-QAM in both TDMA and FDMA.  It can be observed that, our proposed method has significant BER gain over TDMA and FDMA methods, which confirms the effectiveness of the NOMA scheme.  From the simulation results, we also find that FDMA has a smaller BER compared with TDMA. This is because the variance of the effective noise is smaller than that of TDMA. As the rotation based method uses the PSK constellation, which is not spectrally efficient, it has the worst BER performance. Then, in Fig.\,\ref{fig:comparison64qamcase}, the average BER of all the cases are plotted against the SNR where 64-QAM is used by each user for the proposed NOMA and 64-PSK is used by CR based NOMA  while 4096-QAM constellations are used by TDMA and FDMA methods.  We also simulate another two cases with unequal channel variances. Specifically, the average BER of both receivers for all the considered methods is plotted in Fig.\,\ref{fig:weakinterference}, wherein the variances of three channels are set as $(\delta_{11}^2, \delta_{21}^2, \delta_{22}^2)=(1,1/4,1)$. The case with the variances of three channels $(\delta_{11}^2, \delta_{21}^2, \delta_{22}^2)=(1,4,1)$ is provided in Fig.\,\ref{fig:verystronginterference}. Similar observations can be seen as the previous case with equal channel variances, which further verifies the effectiveness of our proposed NOMA design.

\section{Conclusions}
In this paper, we developed a novel and practical design framework for the non-orthogonal multiple access (NOMA) in a classical two-transmitter two-receiver Z-channel with widely-used quadrature amplitude modulation (QAM) and max-min user fairness. Specifically, we formulated a max-min optimization problem to jointly optimize the scaling factors at both transmitters to maximize the smaller minimum Euclidean distance among the two resulting signal constellations at both receivers, subject to the individual average power constraint at each transmitter. The formulated mixed continuous-discrete problem was successfully resolved in compact closed-form by strategically applying the Farey sequences and their unique properties. Simulation results verified the correctness of our analytical derivations and showed that the proposed NOMA design significantly outperforms the existing orthogonal multiple access and NOMA schemes, especially at high signal-to-noise ratio. Furthermore, the performance gap of the proposed scheme over its existing counterparts can be further enlarged when the size of constellations used at the transmitter side becomes larger.
\section*{Appendix}
\begin{appendices}
\subsection{Proof of Proposition\,\ref{prop:mindistant}}\label{appendix:propfareyint}	
We first divide the feasible region $\mathbb Z_{K}^2 \setminus\{(0,0)\}$ into four subsets given by $\mathbb S_{K,1}^2=\{(m,n): m,n \in \mathbb N_{K}, (m,n) \neq (0,0)\}$, $\mathbb S_{K,2}^2=\{(-m,n): m,n \in \mathbb N_{K}, (m,n) \neq (0,0)\}$, $\mathbb S_{K,3}^2=\{(m,-n): m,n \in \mathbb N_{K},  (m,n) \neq (0,0)\}$, $\mathbb S_{K,4}^2=\{(-m,-n): m,n \in \mathbb N_{K}, (m,n) \neq (0,0)\}$. Then clearly we have $\mathbb Z_{K}^2 \setminus \{(0,0)\}  = \cup_{k=1}^4 \mathbb S_{K,k}^2$. For $(-m,n) \in \mathbb S_{K,2}^2$ with $m,n\ge 0$,  we can always find $(m,n) \in \mathbb S_{K,1}^2$ such that
$d_1(m,n)=\big||h_{11}|w_1 n -  |h_{21}| w_2 m\big| \le \big||h_{11}|w_1 n -  |h_{21}| w_2 (-m)\big|= d_1(-m,n)$. Hence,
$\min_{(m,n)\in \mathbb{S}_{K,1}^2}~d_1(m,n)\le \min_{(m,n)\in  \mathbb{S}_{K,2}^2}~d_1(m, n)$.
By a similar argument on $\mathbb{S}_{K,3}^2$ and $\mathbb{S}_{K,4}^2$, it follows that
$\min_{(m,n) \in \mathbb S_{K,1}^2}~d_1(m,n)=\min_{(m,n) \in \mathbb Z_{K,1}^2 \setminus\{(0,0)\} }~d_1(m,n)$.

Then, we further divide $\mathbb S_{K,1}^2$ into $\mathbb{F}^2_{K}$ and $\mathbb S_{K,1}^2 \setminus \mathbb{F}^2_{K}$.
In what follows, we will show that
$\min_{\quad (m,n)\in \mathbb{F}_{K}^2 }~d_1(m,n)=\min_{\quad (m,n)\in \mathbb S_{K,1}^2 }~d_1(m,n)$.
This can be proved by contradiction. Suppose that
$\min_{\quad (m,n) \in \mathbb S_1^2\setminus \mathbb{F}_{K}^2 }~d_1(m,n)=\min_{\quad (m,n) \in \mathbb S_{K,1}^2 }~d_1(m,n)$,
where the minimum is achieved by $(m^*, n^*) \in \mathbb S_{K,1}^2 \setminus \mathbb{F}_{K}^2$ such that $\langle m^*,n^*\rangle=\ell>1$ by the definition of Farey sequences. Then, we can find $(\frac{m^*}{\ell},\frac{n^*}{\ell}) \in \mathbb{F}_{K}^2$ such that $d_1(\frac{m^*}{\ell},\frac{n^*}{\ell})=\frac{1}{\ell} d_1(m^*,n^*)<d_1(m^*,n^*)$, which contradicts the assumption.  This completes the proof.~\hfill$\square$

\subsection{Proof of Proposition\,\ref{prop:basicfareyextended}}\label{appendix:fareydiv}
Recall that $d_1(m,n) =\big||h_{11}|w_1 n-|h_{21}| w_2 m\big|$. Therefore, for $\frac{|h_{21}| w_2}{|h_{11}|w_1}  \in \big(\frac{n_1}{m_1},\frac{n_2}{m_2}\big)$, we have:
$d_1(m_1, n_1) -d_1(m_2, n_2)
=  (m_1+m_2) |h_{11}|w_1 \big(\frac{|h_{21}| w_2}{ |h_{11}|w_1} - \frac{n_1+n_2}{ m_1+m_2}\big)$.
The results presented in the proposition can be readily obtained, and we complete the proof.

\hfill$\square$

\subsection{Proof of of Proposition\,\ref{prop:worstcase}}\label{appendix:worstcase}
We first consider the case $\frac{|h_{21}| w_2}{|h_{11}|w_1} \in (\frac{n_2}{m_2},\frac{n_2+n_3}{m_2+m_3})$. By using Proposition~\ref{prop:basicfareyextended}, we have $d_1(m_2, n_2)<d_1(m_3, n_3)$.
Then, with the help of Property~\ref{prop:basicFareyprop} and~\ref{prop:intervalpart}, we have $\frac{|h_{21}| w_2}{|h_{11}|w_1} \in (\frac{n_2}{m_2}, \frac{n_2+n_3}{m_2+m_3})\subset  (\frac{n_2}{m_2}, \frac{n_3}{m_3}) \subseteq (\frac{n_2}{m_2}, \frac{n_2+n_4}{m_2+m_4})$ and using Property~\ref{prop:basicfareyextended} again, we have $d_1(m_2, n_2)\le d_1(m_4, n_4)$.
On the other hand, by Property~\ref{prop:basicFareyprop} and~\ref{prop:intervalpart}, we have $\frac{|h_{21}| w_2}{|h_{11}|w_1} \in (\frac{n_2}{m_2}, \frac{n_2+n_3}{m_2+m_3})\subset  (\frac{n_2}{m_2}, \frac{n_3}{m_3}) \subseteq (\frac{n_1+n_3}{m_1+m_3}, \frac{n_3}{m_3})$ and using Property~\ref{prop:basicfareyextended} again, we have $d_1(m_3, n_3)\le d_1(m_1, n_1)$.

As $\frac{n_1}{m_1}$ and $\frac{n_4}{m_4}$ are randomly picked entry in $\mathfrak{S}_K$, or equivalently $(m_1, n_1)$ and $(m_4, n_4)$ are randomly picked in $\mathbb F_{M-1}^2 \setminus \{(m_2, n_2),(m_3,n_3)\}$, this proves that for $\frac{|h_{21}| w_2}{|h_{11}|w_1} \in (\frac{n_2}{m_2},\frac{n_2+n_3}{m_2+m_3})$,
$\min_{(m,n) \in \mathbb F_{M-1}^2}~d_1(m,n)=d_1(m_2, n_2)= |h_{21}| w_2 m_2-|h_{11}|w_1 n_2$.
The other case can also be proved by a similar argument and hence omitted.
We completes the proof.
\hfill$\Box$

\subsection{Proof of Proposition~\ref{prop:monotonic}}\label{appendix:monotonic}
First of all, we calculate the following difference
$\frac{a_k+a_{k+1}}{b_k+b_{k+1}} -\frac{a_{k+1}}{b_{k+1}}-\frac{|h_{22}| }{b_{k+1}|h_{21}|}
=\frac{|h_{21}| - |h_{22}|(b_k+b_{k+1})}{(b_k+b_{k+1})b_{k+1} |h_{21}|}$.
Hence, if $\frac{|h_{21}|}{|h_{22}|} \ge b_k+b_{k+1}$, then $\frac{a_{k+1}}{b_{k+1}}+\frac{|h_{22}| }{b_{k+1}|h_{21}|} \le \frac{a_k+a_{k+1}}{b_k+b_{k+1}}$.	
Similarly, we have $\frac{a_{k}}{b_{k}}-\frac{|h_{22}| }{b_{k}|h_{21}|} -\frac{a_k+a_{k+1}}{b_k+b_{k+1}}
=\frac{|h_{21}| - |h_{22}|(b_k+b_{k+1}) }{b_k (b_k+b_{k+1}) |h_{21}|}$.	
As a result, if $\frac{|h_{21}|}{|h_{22}|} \ge b_k+b_{k+1}$, then $ \frac{a_k+a_{k+1}}{b_k+b_{k+1}}\le \frac{a_{k}}{b_{k}}-\frac{|h_{22}| }{b_{k}|h_{21}|} $.	
This completes the proof.\hfill$\Box$	
\subsection{Proof of Lemma\,\ref{thm:subinterval}} \label{appendix:theorem1}
According to proposition~\ref{prop:worstcase} and notice that $\big(\frac{b_k}{a_k},\frac{b_{k+1}}{a_{k+1}}\big)=\big(\frac{b_k}{a_k}, \frac{b_k+b_{k+1}}{a_k+a_{k+1}}\big)\cup \big(\frac{b_k+b_{k+1}}{a_k+a_{k+1}},\frac{b_{k+1}}{a_{k+1}} \big)$, problem in~\eqref{eqn:maxminbyinterval} can be further divided into the following two sub-problems and the overall solution is the maximum value of the two problems:
\begin{problem}[Sub-problem 1]\label{pbm:subpbm1}
	We aim to solve the following optimization problem:
	\begin{subequations}
		\begin{align}
			&g_1\Big(\frac{b_k}{a_k}, \frac{b_{k+1}}{a_{k+1}}\Big)=\max_{w_1, w_2}~\min~\{|h_{11}| w_1 b_{k+1} \nonumber\\
			&\qquad\qquad\qquad\qquad\qquad - |h_{21}|w_2 a_{k+1},|h_{22}|w_2 \}\\
			&{\rm s.t.~}\frac{b_k+b_{k+1}}{a_k+a_{k+1}} \le \frac{|h_{21}|w_2 }{|h_{11}| w_1}\le \frac{b_{k+1}}{a_{k+1}},\\
			&\quad 0< w_1 \le \sqrt{\frac{3 p_1}{M^2-1}}, 0< w_2 \le \sqrt{\frac{3 p_2}{M^2-1}}.
		\end{align}
		\hfill$\blacksquare$
	\end{subequations}	
\end{problem}
\begin{problem}[Sub-problem 2]\label{pbm:subpbm2}
	The optimization problem is stated as follows:
	\begin{subequations}
		\begin{align}
			&g_2\Big(\frac{b_k}{a_k}, \frac{b_{k+1}}{a_{k+1}}\Big)=\max_{w_1, w_2}\min\{|h_{21}|w_2 a_k \nonumber\\
			&\qquad\qquad\qquad\qquad\qquad - |h_{11}| w_1 b_k, |h_{22}|w_2 \}\\
			&{\rm s.t.~}\frac{b_k}{a_k}\le \frac{|h_{21}|w_2 }{|h_{11}| w_1} < \frac{b_k+b_{k+1}}{a_k+a_{k+1}},\\
			&\quad 0< w_1 \le \sqrt{\frac{3 p_1}{M^2-1}},~0< w_2 \le \sqrt{\frac{3 p_2}{M^2-1}}.
		\end{align}
		\hfill$\blacksquare$
	\end{subequations}	
\end{problem}

We first consider Sub-problem 1 and Problem~\ref{pbm:subpbm1}, which can be divided into the following two case:	
\begin{enumerate}
	\item Case 1: Receiver 1 has smaller Euclidean distance.
	\begin{subequations}
		\begin{align}
			&g_{11}\Big(\frac{b_k}{a_k}, \frac{b_{k+1}}{a_{k+1}}\Big)=\max_{w_1, w_2}~  b_{k+1} |h_{11}| w_1 -a_{k+1} |h_{21}|w_2\nonumber\\
			&{\rm s.t.~} b_{k+1}|h_{11}| w_1  -a_{k+1} |h_{21}|w_2  \le |h_{22}|w_2,\label{const:prob4case1const1}\\
			&\quad \frac{b_k+b_{k+1}}{a_k+a_{k+1}}\le \frac{|h_{21}|w_2 }{|h_{11}| w_1}\le\frac{b_{k+1}}{a_{k+1}},\label{const:prob4case1const2}\\
			&\quad 0< w_1 \le \sqrt{\frac{3 p_1}{M^2-1}}, 0< w_2 \le \sqrt{\frac{3 p_2}{M^2-1}}.
		\end{align}
	\end{subequations}	
	\item Case 2: Receiver 2 has smaller Euclidean distance.
	\begin{subequations}
		\begin{align}
			&g_{12}\Big(\frac{b_k}{a_k}, \frac{b_{k+1}}{a_{k+1}}\Big)=\max_{w_1, w_2}~|h_{22}|w_2 \nonumber\\
			&{\rm s.t.~} |h_{22}|w_2<b_{k+1} |h_{11}| w_1-a_{k+1}|h_{21}|w_2,\label{const:prob4case2const1}\\
			&\quad \frac{b_k+b_{k+1}}{a_k+a_{k+1}}\le \frac{|h_{21}|w_2 }{|h_{11}| w_1}<\frac{b_{k+1}}{a_{k+1}},\label{const:prob4case2const2}\\
			&\quad 0< w_1 \le \sqrt{\frac{3 p_1}{M^2-1}}, 0< w_2 \le \sqrt{\frac{3 p_2}{M^2-1}}.
		\end{align}
	\end{subequations}	
\end{enumerate}

For {\bf{Case-1 of Sub-problem 1}}, constraint~\eqref{const:prob4case1const1} is equivalent to $\frac{b_{k+1} |h_{11}| }{a_{k+1}|h_{21}|+|h_{22}|}w_1 \le w_2$ and  constraint~\eqref{const:prob4case1const2} means $\frac{(b_k+b_{k+1}) |h_{11}|}{(a_k+a_{k+1})|h_{21}|} w_1\le w_2\le\frac{b_{k+1} |h_{11}|}{a_{k+1} |h_{21}|} w_1$.
Also we notice that if $\frac{|h_{21}|}{|h_{22}|} \ge b_k+b_{k+1}$ then $\frac{(b_k+b_{k+1})|h_{11}|}{(a_k+a_{k+1})|h_{21}|}w_1  \le \frac{ b_{k+1} |h_{11}| }{a_{k+1}|h_{21}|+|h_{22}|} w_1$.
Hence, the optimization problem can be further divided into two cases:
\begin{enumerate}
	\item If $\frac{|h_{21}|}{|h_{22}|} \le b_k +b_{k+1}$, then
	\begin{subequations}
		\begin{align}
			&g_{11}\Big(\frac{b_k}{a_k}, \frac{b_{k+1}}{a_{k+1}}\Big)=\max_{w_1, w_2}~ b_{k+1} |h_{11}| w_1-a_{k+1}|h_{21}|w_2\nonumber\\
			&{\rm s.t.~}
			\frac{(b_k+b_{k+1}) |h_{11}|}{(a_k+a_{k+1})|h_{21}|} w_1\le w_2\le\frac{b_{k+1} |h_{11}|}{a_{k+1} |h_{21}|},\\
			&\quad 0< w_1 \le \sqrt{\frac{3 p_1}{M^2-1}}, 0< w_2 \le \sqrt{\frac{3 p_2}{M^2-1}}\label{const:prob4case1const3}.
		\end{align}
	\end{subequations}
	We let $w_2=\frac{(b_k+b_{k+1}) |h_{11}|}{(a_k+a_{k+1})|h_{21}|} w_1$, then the objective function is $\frac{|h_{11}| w_1}{a_k+a_{k+1}}$.
	In this case, \eqref{const:prob4case1const3} is equivalent to $w_1 \le \frac{(a_k+a_{k+1})|h_{21}|}{(b_k+b_{k+1}) |h_{11}|}  \sqrt{\frac{3 p_2}{M^2-1}}$ and $w_1 \le \sqrt{\frac{3 p_1}{M^2-1}}$ and hence, we have $g_{11}\big(\frac{b_k}{a_k}, \frac{b_{k+1}}{a_{k+1}}\big)=\min \Big\{
	\frac{|h_{11}|}{a_k+a_{k+1}}\frac{(a_k+a_{k+1})|h_{21}|}{(b_k+b_{k+1}) |h_{11}|}  \sqrt{\frac{3 p_2}{M^2-1}}, \frac{|h_{11}|}{a_k+a_{k+1}} \sqrt{\frac{3 p_1}{M^2-1}}\Big\}$. As a consequence, we have
	\begin{subequations}
		\begin{align*}
			&g_{11}\Big(\frac{b_k}{a_k}, \frac{b_{k+1}}{a_{k+1}}\Big)\nonumber \\ 
			&=\begin{cases}\frac{|h_{21}|}{b_k+b_{k+1}} \sqrt{\frac{3 p_2}{M^2-1}},{\rm~where~}\\
			(w_1, w_2)= \big(\frac{(a_k+a_{k+1})|h_{21}|}{(b_k+b_{k+1}) |h_{11}|}  \sqrt{\frac{3 p_2}{M^2-1}}, \sqrt{\frac{3 p_2}{M^2-1}}\big),\\
			\qquad{\rm~if~}\frac{|h_{11}|}{|h_{21}|}\ge \frac{ \sqrt{p_2}(a_k+ a_{k+1})}{\sqrt{p_1} (b_k+b_{k+1})};\\
				\frac{|h_{11}|}{a_k+ a_{k+1}}\sqrt{\frac{3 p_1}{M^2-1}},
				{\rm~where~} \\
				(w_1, w_2)=\big(\sqrt{\frac{3 p_1}{M^2-1}},\frac{(b_k+b_{k+1}) |h_{11}|}{(a_k+a_{k+1})|h_{21}|}\sqrt{\frac{3 p_1}{M^2-1}}\big), \\
				\qquad{\rm~if~}\frac{|h_{11}|}{|h_{21}|}< \frac{ \sqrt{p_2}(a_k+ a_{k+1})}{\sqrt{p_1} (b_k+b_{k+1})},
			\end{cases}
		\end{align*}
	\end{subequations}
	\item If $\frac{|h_{21}|}{|h_{22}|} > b_k +b_{k+1}$, then
	\begin{subequations}
		\begin{align}
			&g_{11}\Big(\frac{b_k}{a_k}, \frac{b_{k+1}}{a_{k+1}}\Big)=\max_{w_1, w_2}~ b_{k+1} |h_{11}| w_1-a_{k+1}|h_{21}|w_2 \nonumber\\
			&{\rm s.t.}
			\frac{b_{k+1} |h_{11}| }{a_{k+1}|h_{21}|+|h_{22}|}w_1\le w_2\le\frac{b_{k+1} |h_{11}|}{a_{k+1} |h_{21}|} w_1,\\
			&\quad 0< w_1 \le \sqrt{\frac{3 p_1}{M^2-1}}, 0< w_2 \le \sqrt{\frac{3 p_2}{M^2-1}}\label{const:prob4case1const4}.
		\end{align}
	\end{subequations}
\end{enumerate}
We first notice that $\frac{b_{k+1} |h_{11}| }{a_{k+1}|h_{21}|+|h_{22}|}w_1<\frac{b_{k+1} |h_{11}|}{a_{k+1} |h_{21}|} w_1$ and hence the problem is always feasible.
By letting $w_2=\frac{b_{k+1} |h_{11}| }{a_{k+1}|h_{21}|+|h_{22}|}w_1$, the objective function can be written by $\frac{b_{k+1} |h_{11}||h_{22}|}{a_{k+1}|h_{21}|+|h_{22}|}w_1$.
In this case, the constraints in~\eqref{const:prob4case1const4} are equivalent to $w_1 \le \sqrt{\frac{3 p_1}{M^2-1}}$ and $w_1 \le \frac{a_{k+1}|h_{21}|+|h_{22}|}{b_{k+1} |h_{11}| }  \sqrt{\frac{3 p_2}{M^2-1}}$.
Therefore, the solution is
\begin{subequations}
	\begin{align*}
		&g_{11}\Big(\frac{b_k}{a_k}, \frac{b_{k+1}}{a_{k+1}}\Big) \nonumber\\
		&=\begin{cases}\frac{b_{k+1}|h_{11}||h_{22}|}{a_{k+1}|h_{21}|+|h_{22}|}\sqrt{\frac{3 p_1}{M^2-1}},
			{\rm~where~} \\
			(w_1, w_2)=\big(\sqrt{\frac{3 p_1}{M_1^2-1}}, \frac{b_{k+1}|h_{11}|}{a_{k+1}|h_{21}|+|h_{22}|}\sqrt{\frac{3 p_1}{M^2-1}}\big),
			\\ \qquad{\rm~if~}\frac{|h_{11}|}{|h_{21}|}\le \frac{\sqrt{p_2}}{\sqrt{p_1}}\big(\frac{a_{k+1}}{b_{k+1}}
			+\frac{|h_{22}| }{b_{k+1}|h_{21}|}\big);\\
			|h_{22}|\sqrt{\frac{3 p_2}{M_2^2-1}}, {\rm~where~} \\
			(w_1, w_2)=\big(\frac{a_{k+1}|h_{21}|+|h_{22}|}{b_{k+1} |h_{11}| }  \sqrt{\frac{3 p_2}{M^2-1}}, \sqrt{\frac{3 p_2}{M^2-1}}\big),\\
			 \qquad{\rm~if~} \frac{|h_{11}|}{|h_{21}|}> \frac{\sqrt{p_2}}{\sqrt{p_1}} \big(\frac{a_{k+1}}{b_{k+1}}
			+\frac{|h_{22}| }{b_{k+1}|h_{21}|}\big).
		\end{cases}
	\end{align*}
\end{subequations}
For {\bf{Case-2 of Sub-problem 1}},
constraint~\eqref{const:prob4case2const1} is equivalent to $w_2<\frac{ b_{k+1} |h_{11}| }{a_{k+1}|h_{21}|+|h_{22}|} w_1$ and constraint~\eqref{const:prob4case2const2} implies $\frac{(b_k+b_{k+1})|h_{11}|}{(a_k+a_{k+1})|h_{21}|}w_1\le w_2 \le  \frac{b_{k+1} |h_{11}|}{a_{k+1}|h_{21}|} w_1$. By noticing that $\frac{ b_{k+1} |h_{11}| }{a_{k+1}|h_{21}|+|h_{22}|}< \frac{b_{k+1} |h_{11}|}{a_{k+1}|h_{21}|}$,
the problem is equivalent to
\begin{subequations}
	\begin{align}
		&g_{12}\Big(\frac{b_k}{a_k}, \frac{b_{k+1}}{a_{k+1}}\Big)=\max_{w_1, w_2}~ |h_{22}|w_2 \nonumber\\
		&{\rm  s.t.~}\frac{(b_k+b_{k+1})|h_{11}|}{(a_k+a_{k+1})|h_{21}|}w_1\le w_2 \le \frac{ b_{k+1} |h_{11}| }{a_{k+1}|h_{21}|+|h_{22}|} w_1,\label{const:prob4case2const3}\\
		&\quad 0< w_1 \le \sqrt{\frac{3 p_1}{M^2-1}},0< w_2 \le \sqrt{\frac{3 p_2}{M^2-1}}.
	\end{align}
\end{subequations}
Constraint~\eqref{const:prob4case2const3} is feasible if $\frac{|h_{21}|}{|h_{22}|} \ge b_k+b_{k+1}$.
In this case, the solution is
\begin{subequations}
	\begin{align*}
		&g_{12}\Big(\frac{b_k}{a_k}, \frac{b_{k+1}}{a_{k+1}}\Big) \nonumber\\
		&=\begin{cases}\frac{b_{k+1}|h_{11}||h_{22}|}{a_{k+1}|h_{21}|+|h_{22}|}\sqrt{\frac{3 p_1}{M^2-1}},
			{\rm~where~}\\
			 (w_1, w_2)=\big(\sqrt{\frac{3 p_1}{M^2-1}},\frac{b_{k+1}|h_{11}|}{a_{k+1}|h_{21}|+|h_{22}|}\sqrt{\frac{3 p_1}{M^2-1}}\big),
			\\\qquad {\rm~if~}\frac{|h_{11}|}{|h_{21}|}\le \frac{\sqrt{p_2}}{\sqrt{p_1}}\big(\frac{a_{k+1}}{b_{k+1}}
			+\frac{|h_{22}| }{b_{k+1}|h_{21}|}\big);\\
			|h_{22}|\sqrt{\frac{3 p_2}{M_2^2-1}},
			{\rm~where~}\\
			 (w_1, w_2)=\big(\frac{a_{k+1}|h_{21}|+|h_{22}|}{ b_{k+1} |h_{11}| }\sqrt{\frac{3 p_2}{M_2^2-1}},\sqrt{\frac{3 p_2}{M_2^2-1}}\big),
			\\\qquad {\rm~if~}\frac{|h_{11}|}{|h_{21}|}> \frac{\sqrt{p_2}}{\sqrt{p_1}} \big(\frac{a_{k+1}}{b_{k+1}}
			+\frac{|h_{22}| }{b_{k+1}|h_{21}|}\big).
		\end{cases}
	\end{align*}
\end{subequations}

The solution of Sub-problem 2 can be attained in a similar fashion as Sub-problem 1, and hence is omitted.
Then, for the subinterval division $\frac{|h_{21}|w_2 }{|h_{11}| w_1} \in \big(\frac{b_k}{a_k}, \frac{b_{k+1}}{a_{k+1}}\big)$, {\bf{the solution to both Sub-problems can be summarized as follows:}}
\begin{enumerate}
	\item Scenario 1, $\frac{b_k+b_{k+1}}{a_k+a_{k+1}}\le \frac{|h_{21}|w_2 }{|h_{11}| w_1}\le\frac{b_{k+1}}{a_{k+1}}$:
	\begin{enumerate}
		\item  If $\frac{|h_{21}|}{|h_{22}|} \le b_k +b_{k+1}$
		\begin{subequations}
			\begin{align*}
				&g_{11}\Big(\frac{b_k}{a_k}, \frac{b_{k+1}}{a_{k+1}}\Big)\\
				&=\begin{cases}\frac{|h_{21}|}{b_k+b_{k+1}} \sqrt{\frac{3 p_2}{M^2-1}},{\rm~where~}\\
				  (w_1, w_2) = \big(\frac{(a_k+a_{k+1})|h_{21}|}{(b_k+b_{k+1}) |h_{11}|}  \sqrt{\frac{3 p_2}{M^2-1}},\sqrt{\frac{3 p_2}{M^2-1}}\big),\\
				\qquad{\rm~if~}\frac{|h_{11}|}{|h_{21}|}\ge \frac{ \sqrt{p_2}(a_k+ a_{k+1})}{\sqrt{p_1} (b_k+b_{k+1})};\\
					\frac{|h_{11}|}{a_k+ a_{k+1}}\sqrt{\frac{3 p_1}{M^2-1}},{\rm~where~}\\
					 (w_1, w_2)=\big(\sqrt{\frac{3 p_1}{M^2-1}}, \frac{(b_k+b_{k+1}) |h_{11}|}{(a_k+a_{k+1})|h_{21}|}\sqrt{\frac{3 p_1}{M^2-1}}\big), \\
					\qquad {\rm~if~}\frac{|h_{11}|}{|h_{21}|}< \frac{ \sqrt{p_2}(a_k+ a_{k+1})}{\sqrt{p_1} (b_k+b_{k+1})},
				\end{cases}
			\end{align*}
		\end{subequations}
		\item  If $\frac{|h_{21}|}{|h_{22}|} > b_k +b_{k+1}$
		\begin{subequations}
			\begin{align*}
				&g_{11}\Big(\frac{b_k}{a_k}, \frac{b_{k+1}}{a_{k+1}}\Big)\\
				 &=\begin{cases}\frac{b_{k+1}|h_{11}||h_{22}|}{a_{k+1}|h_{21}|+|h_{22}|}\sqrt{\frac{3 p_1}{M^2-1}},
					{\rm~where~} (w_1, w_2)=\\
					\big(\sqrt{\frac{3 p_1}{M^2-1}},\frac{b_{k+1}|h_{11}|}{a_{k+1}|h_{21}|+|h_{22}|}\sqrt{\frac{3 p_1}{M^2-1}}\big),
					\\\qquad {\rm~if~}\frac{|h_{11}|}{|h_{21}|}\le \frac{\sqrt{p_2}}{\sqrt{p_1}}\big(\frac{a_{k+1}}{b_{k+1}}
					+\frac{|h_{22}| }{b_{k+1}|h_{21}|}\big);\\
					|h_{22}|\sqrt{\frac{3 p_2}{M^2-1}}, {\rm~where~}(w_1, w_2)=\\
					 \big(\frac{a_{k+1}|h_{21}|+|h_{22}|}{b_{k+1} |h_{11}| }  \sqrt{\frac{3 p_2}{M^2-1}}, \sqrt{\frac{3 p_2}{M^2-1}}\big),\\
					\qquad {\rm~if~} \frac{|h_{11}|}{|h_{21}|}> \frac{\sqrt{p_2}}{\sqrt{p_1}} \big(\frac{a_{k+1}}{b_{k+1}}
					+\frac{|h_{22}| }{b_{k+1}|h_{21}|}\big).
				\end{cases}
			\end{align*}
		\end{subequations}
	\end{enumerate}
	\item Scenario 2, $\frac{b_k}{a_k}\le \frac{|h_{21}|w_2 }{|h_{11}| w_1}\le \frac{b_k+b_{k+1}}{a_k+a_{k+1}}$:
	\begin{enumerate}
		\item Case 1: Receiver 1 has smaller Euclidean distance.
		\begin{enumerate}
			\item If $\frac{|h_{21}|}{|h_{22}|}\le b_k+b_{k+1}$, the solution is
			\begin{subequations}
				\begin{align*}
					&g_{21}\Big(\frac{b_k}{a_k}, \frac{b_{k+1}}{a_{k+1}}\Big)\\
					&=\begin{cases}\frac{|h_{21}|}{b_k+b_{k+1}}\sqrt{\frac{3 p_2}{M^2-1}},{~\rm where~}(w_1, w_2)=\\
						\big(\frac{(a_k+a_{k+1})|h_{21}|}{(b_k+b_{k+1})|h_{11}|}\sqrt{\frac{3 p_2}{M^2-1}},\sqrt{\frac{3 p_2}{M^2-1}}\big),	\\
						\qquad {\rm~if~} \frac{|h_{11}|}{|h_{21}|}\ge \frac{\sqrt{p_2}(a_k+a_{k+1})}{\sqrt{p_1}(b_k+b_{k+1})};
						\\
						\frac{|h_{11}|}{a_k+a_{k+1}}\sqrt{\frac{3 p_1}{M^2-1}}, {\rm~where~}(w_1, w_2)=\\
						\big(\sqrt{\frac{3 p_1}{M^2-1}}, \frac{(b_k+b_{k+1})|h_{11}|}{(a_k+a_{k+1})|h_{21}|} \sqrt{\frac{3 p_1}{M^2-1}}\big),\\
						\qquad {\rm~if~}\frac{|h_{11}|}{|h_{21}|}< \frac{\sqrt{p_2}(a_k+a_{k+1})}{\sqrt{p_1}(b_k+b_{k+1})}.
					\end{cases}
				\end{align*}
			\end{subequations}
			\item If $\frac{|h_{21}|}{|h_{22}|}\ge b_k+b_{k+1}$,
			\begin{subequations}
				\begin{align*}
					&g_{21}\Big(\frac{b_k}{a_k}, \frac{b_{k+1}}{a_{k+1}}\Big)\\
					&=\begin{cases}|h_{22}|\sqrt{\frac{3 p_2}{M^2-1}}, {\rm~where~}(w_1, w_2)=\\
					 \big(\frac{|h_{21}| a_k-|h_{22}|}{b_k |h_{11}|}\sqrt{\frac{3 p_2}{M^2-1}}, \sqrt{\frac{3 p_2}{M^2-1}}\big),\\
						\qquad {\rm~if~}\frac{|h_{11}|}{|h_{21}|}\ge \frac{\sqrt{p_2} }{\sqrt{p_1}}\big( \frac{a_k}{b_k} -\frac{|h_{22}|}{|h_{21}|b_k}\big);\\
						\frac{b_k |h_{11}||h_{22}| }{|h_{21}| a_k-|h_{22}|}\sqrt{\frac{3 p_1}{M^2-1}}, {\rm~where~}(w_1, w_2)=\\
						 \big(\sqrt{\frac{3 p_1}{M^2-1}}, \frac{b_k |h_{11}|}{|h_{21}| a_k-|h_{22}|}\sqrt{\frac{3 p_1}{M^2-1}}\big),\\
						\qquad {\rm~if~}\frac{|h_{11}|}{|h_{21}|}<\frac{\sqrt{p_2} }{\sqrt{p_1}}\big( \frac{a_k}{b_k} -\frac{|h_{22}|}{|h_{21}|b_k}\big).
					\end{cases}
				\end{align*}
			\end{subequations}
		\end{enumerate}
		\item Case 2: Receiver 2 has smaller Euclidean distance.
		\begin{enumerate}
			\item If $\frac{|h_{21}|}{|h_{22}|} \ge b_k +b_{k+1}$, the solution is
			\begin{subequations}
				\begin{align*}
					&g_{22}\Big(\frac{b_k}{a_k}, \frac{b_{k+1}}{a_{k+1}}\Big)\\
					 &=\begin{cases}\frac{(b_k+b_{k+1})|h_{11}||h_{22}|}{(a_k+a_{k+1})|h_{21}|}\sqrt{\frac{3 p_1}{M^2-1}}, {\rm ~where~}(w_1, w_2)\\= \big(\sqrt{\frac{3 p_1}{M^2-1}}, \frac{(b_k+b_{k+1})|h_{11}|}{(a_k+a_{k+1})|h_{21}|}\sqrt{\frac{3 p_1}{M^2-1}}\big),\\
						\quad {\rm~if~}\frac{|h_{11}| }{|h_{21}|}\le \frac{\sqrt{p_2}(a_k+a_{k+1})}{\sqrt{p_1}(b_k+b_{k+1})};\\
						|h_{22}|\sqrt{\frac{3 p_2}{M^2-1}}, {\rm ~where~}(w_1, w_2)\!=\!\\
						 \big(\frac{(a_k+a_{k+1})|h_{21}|}{(b_k+b_{k+1})|h_{11}|}\sqrt{\frac{3 p_2}{M^2-1}},\sqrt{\frac{3 p_2}{M^2-1}}\big),\\
						\qquad {\rm~if~}\frac{|h_{11}| }{|h_{21}|}>\frac{\sqrt{p_2}(a_k+a_{k+1})}{\sqrt{p_1}(b_k+b_{k+1})}.
					\end{cases}
				\end{align*}
			\end{subequations}
		\end{enumerate}
	\end{enumerate}
\end{enumerate}
Now, we aim to combine $g_{21}\big(\frac{b_k}{a_k}, \frac{b_{k+1}}{a_{k+1}}\big)$ and $g_{22}\big(\frac{b_k}{a_k}, \frac{b_{k+1}}{a_{k+1}}\big)$.
By Proposition~\ref{prop:monotonic}, for $\frac{|h_{21}|}{|h_{22}|} \ge b_k +b_{k+1}$, we have $\frac{a_k}{b_k} -\frac{|h_{22}|}{|h_{21}|b_k}\ge \frac{a_k+a_{k+1}}{b_k+b_{k+1}}$.
Also, for $\frac{|h_{11}|}{|h_{21}|} \le \frac{\sqrt{p_2} } {\sqrt{p_1}} \big(\frac{a_k}{b_k}-\frac{|h_{22}|}{b_k |h_{21}|}\big)$, we have $\sqrt{\frac{3 p_2}{M^2-1}}\ge \frac{b_k |h_{11}| }{|h_{21}| a_k-|h_{22}|}\sqrt{\frac{3 p_1}{M^2-1}}$. In addition, for $\frac{|h_{21}|}{|h_{22}|} \ge b_k+b_{k+1}$, we have $\frac{(b_k+b_{k+1})}{(a_k+a_{k+1})|h_{21}|}\ge \frac{b_k }{|h_{21}| a_k-|h_{22}|}$.
Hence, we have $g_{22}\big(\frac{b_k}{a_k}, \frac{b_{k+1}}{a_{k+1}}\big) \ge g_{21}\big(\frac{b_k}{a_k}, \frac{b_{k+1}}{a_{k+1}}\big)$.
We notice that for $\frac{|h_{21}|}{|h_{22}|} \le b_k +b_{k+1}$, $g_{11}\big(\frac{b_k}{a_k}, \frac{b_{k+1}}{a_{k+1}}\big)=g_{21}\big(\frac{b_k}{a_k}, \frac{b_{k+1}}{a_{k+1}}\big)$. Then for $\frac{|h_{21}|}{|h_{22}|} \ge b_k +b_{k+1}$, we combine $g_{11}\big(\frac{b_k}{a_k}, \frac{b_{k+1}}{a_{k+1}}\big)$ and $g_{22}\big(\frac{b_k}{a_k}, \frac{b_{k+1}}{a_{k+1}}\big)$.
By Proposition~\ref{prop:monotonic}, for $\frac{|h_{21}|}{|h_{22}|}\ge b_k+b_{k+1}$, we have $\frac{\sqrt{p_2}}{\sqrt{p_1}}\big(\frac{a_{k+1}}{b_{k+1}}
+\frac{|h_{22}| }{b_{k+1}|h_{21}|}\big) \le \frac{\sqrt{p_2}(a_k+a_{k+1})}{\sqrt{p_1}(b_k+b_{k+1})}$.
Also, for $\frac{|h_{21}|}{ |h_{22}|}  \ge  b_k +b_{k+1}$,  we attain $\frac{b_{k+1}|h_{11}||h_{22}|}{a_{k+1}|h_{21}|+|h_{22}|}  \sqrt{\frac{3 p_1}{M^2-1}}\ge\frac{(b_k+b_{k+1})|h_{11}||h_{22}|}{(a_k+a_{k+1})|h_{21}|}\sqrt{\frac{3 p_1}{M^2-1}}$. In addition, for $\frac{|h_{11}|}{|h_{21}|}\le \frac{\sqrt{p_2} (a_k+a_{k+1})}{\sqrt{p_1}(b_k+b_{k+1})}$, we have $|h_{22}|\sqrt{\frac{3 p_2}{M^2-1}} \ge \frac{(b_k+b_{k+1})|h_{11}||h_{22}|}{(a_k+a_{k+1})|h_{21}|}\sqrt{\frac{3 p_1}{M^2-1}}$. In conclusion, for $\frac{|h_{21}|}{|h_{22}|} \ge b_k +b_{k+1}$, $g_{11}\big(\frac{b_k}{a_k}, \frac{b_{k+1}}{a_{k+1}}\big) \ge g_{22}\big(\frac{b_k}{a_k}, \frac{b_{k+1}}{a_{k+1}}\big)$.

With the above discussion, we have the result in Lemma~\ref{thm:subinterval} and we complete the proof.\hfill$\Box$

\subsection{Proof of Theorem~\ref{thm:weakint}}\label{appendix:theoremweakint}
For the weak cross link, we have $\frac{|h_{21}|}{|h_{22}|} \le 1 \le b_k +b_{k+1}$ for $k=1, \ldots, |\mathcal{S}_{M-1}|$.
\begin{enumerate}
	\item If $\frac{|h_{11}|}{|h_{21}|} \le \frac{ \sqrt{p_2} }{M \sqrt{p_1}}$, we have $\frac{|h_{11}|}{|h_{21}| } \le  \frac{\sqrt{p_2} (a_{k} +a_{k+1}) }{\sqrt{p_1}(b_{k} +b_{k+1})}$, and  then by Lemma~\ref{thm:subinterval}, we attain $g\big(\frac{b_{k}}{a_{k}}, \frac{b_{k+1}}{a_{k+1}}\big)=
	\frac{|h_{11}|}{a_{k} + a_{k+1}}\sqrt{\frac{3 p_1}{M^2-1}} ~{\rm for~} k=1, \ldots, |\mathcal{S}_{M-1}|$.
	Also, note that $|\mathcal{S}_{M-1}| =\arg \min_k~\{(a_{1}+ a_{2}), \ldots, (a_{ |\mathcal{S}_{M-1}|}+ a_{ |\mathcal{S}_{M-1}|+1})\}$, then $\max~\Big\{
		\frac{|h_{11}|}{a_{1}+ a_{2}}\sqrt{\frac{3 p_1}{M^2-1}},\cdots, \frac{|h_{11}|}{a_{ |\mathcal{S}_{M-1}|}+ a_{ |\mathcal{S}_{M-1}|+1}}\sqrt{\frac{3 p_1}{M^2-1}}
		\Big\}
		=|h_{11}|\sqrt{\frac{3 p_1}{M^2-1}}$,
	with $(w_1^*, w_2^*)=\Big(\sqrt{\frac{3 p_1}{M^2-1}}, \frac{M |h_{11}|}{|h_{21}|}\sqrt{\frac{3 p_1}{M^2-1}}\Big)$.
	
	\item If $\frac{|h_{11}|}{|h_{21}|} \ge \frac{M\sqrt{p_2} }{\sqrt{p_1}}$, we have $\frac{|h_{11}|}{|h_{21}| } \ge \frac{(a_{k} +a_{k+1}) \sqrt{p_2} }{(b_{k} +b_{k+1})\sqrt{p_1}}$ and then, by  using Lemma~\ref{thm:subinterval}, we have $g\big(\frac{b_{k}}{a_{k}}, \frac{b_{k+1}}{a_{k+1}}\big)= \frac{|h_{21}|}{b_{k} + b_{k+1}}\sqrt{\frac{3 p_2}{M_2^2-1}}$ for $k=1, \ldots, |\mathcal{S}_{M-1}|$. Also, note that $1=\arg \min_k~\{(b_{1}+ b_{2}), \ldots, (b_{ |\mathcal{S}_{M-1}|}+ b_{ |\mathcal{S}_{M-1}|+1})\}$.  Then,
	\begin{align*}
		&\max~\Big\{
		\frac{|h_{21}|}{b_{1}+ b_{2}}\sqrt{\frac{3 p_2}{M^2-1}},\cdots, \\
		&\frac{|h_{21}|}{b_{ |\mathcal{S}_{M-1}|}+ b_{ |\mathcal{S}_{M-1}|+1}}\sqrt{\frac{3 p_2}{M^2-1}}
		\Big\}
		=|h_{21}|\sqrt{\frac{3 p_2}{M^2-1}},
	\end{align*}
	with $(w_1^*, w_2^*)=\Big(\frac{M |h_{21}|}{|h_{11}|}\sqrt{\frac{3 p_2}{M^2-1}},\sqrt{\frac{3 p_2}{M^2-1}}\Big)$.
	\item If $\frac{|h_{11}|}{|h_{21}| } \in \Big(\frac{\sqrt{p_2}(a_{\ell_1+1} +a_{\ell_1+2})  }{\sqrt{p_1}(b_{\ell_1+1} +b_{\ell_1+2})}, \frac{ \sqrt{p_2} (a_{\ell_1} +a_{\ell_1+1})}{\sqrt{p_1}(b_{\ell_1} +b_{\ell_1+1})}\Big)$, for $\ell_1=1, \ldots,   |\mathcal{S}_{M-1}|-1$, then, with the help of Lemma~\ref{thm:subinterval}, we have
	\begin{align*}
		&g\Big(\frac{b_{k}}{a_{k}}, \frac{b_{k+1}}{a_{k+1}}\Big)\\
		&=\begin{cases}
			\frac{|h_{11}|}{a_{k} + a_{k+1}}\sqrt{\frac{3 p_1}{M^2-1}} ~{\rm for~}k=1, \ldots, \ell_1, \\
			\frac{|h_{21}|}{b_{k}+b_{k+1}} \sqrt{\frac{3 p_2}{M^2-1}} ~{\rm~for~} k=\ell_1+1, \ldots,|\mathcal{S}_{M-1}|.
		\end{cases}
	\end{align*}
	
	Note that, $\tilde{\ell}_a=\arg \min_k \{(a_1+a_2), \ldots, (a_{\ell_1}+ a_{\ell_1+1}) \}$ and
	$\tilde{\ell}_b=\arg \min_k \{(b_{\ell_1+1}+b_{\ell_1+2}), \ldots, (b_{|\mathcal{S}_{M-1}|}+b_{|\mathcal{S}_{M-1}|+1})\}$, hence we have
\begin{footnotesize}
	\begin{align*}
		&\max~\Big\{
		\frac{|h_{11}|}{a_{1}+ a_{2}}\sqrt{\frac{3 p_1}{M^2-1}},\cdots, \frac{|h_{11}|}{a_{\ell_1}+ a_{\ell_1+1}}\sqrt{\frac{3 p_1}{M^2-1}},\\
        &\!\!\frac{|h_{21}|}{b_{\ell_1+1}+b_{\ell_1+2}} \sqrt{\frac{3 p_2}{M^2-1}}, \ldots,\frac{|h_{21}|}{b_{|\mathcal{S}_{M-1}|}+b_{|\mathcal{S}_{M-1}|+1}} \sqrt{\frac{3 p_2}{M^2-1}}
		\Big\}\\
		&=\max~ \Big\{
		\frac{|h_{11}|}{a_{\tilde{\ell}_a}+ a_{\tilde{\ell}_a+1}}\sqrt{\frac{3 p_1}{M^2-1}},\frac{|h_{21}|}{b_{\tilde{\ell}_b}+b_{\tilde{\ell}_b+1}} \sqrt{\frac{3 p_2}{M^2-1}}\Big\}.
	\end{align*}
\end{footnotesize}
	Therefore, if $\frac{|h_{11}|}{|h_{21}|} \ge \frac{\sqrt{ p_2}(a_{\tilde{\ell}_a}+ a_{\tilde{\ell}_a+1}) }{\sqrt{ p_1}(b_{\tilde{\ell}_b}+b_{\tilde{\ell}_b+1})}$, we have
	$\frac{|h_{11}|}{a_{\tilde{\ell}_a}+ a_{\tilde{\ell}_a+1}}\sqrt{\frac{3 p_1}{M^2-1}}\ge \frac{|h_{21}|}{b_{\tilde{\ell}_b}+b_{\tilde{\ell}_b+1}} \sqrt{\frac{3 p_2}{M^2-1}}$ and hence,
	$(w_1^*, w_2^*)=\Big(\sqrt{\frac{3 p_1}{M^2-1}},\frac{(b_{\tilde{\ell}_a} +b_{\tilde{\ell}_a+1})|h_{11}|}{(a_{\tilde{\ell}_a}+a_{\tilde{\ell}_a+1})|h_{21}|}\sqrt{\frac{3 p_1}{M-1}}\Big)$ and else we have $(w_1^*, w_2^*) = \Big(\frac{(a_{\tilde{\ell}_b} + a_{\tilde{\ell}_b+1})|h_{21}|}{(b_{\tilde{\ell}_b}+b_{\tilde{\ell}_b+1}) |h_{11}|}\sqrt{\frac{3 p_2}{M^2-1}}, \sqrt{\frac{3 p_2}{M^2-1}}\Big)$.
\end{enumerate}

This completes the proof.\hfill$\Box$
%
\subsection{Proof of Theorem~\ref{thm:verystrongint}}\label{appendix:theoremverystrongint}
\begin{enumerate}
	\item If $|h_{11}| \le \frac{ \sqrt{p_2}}{\sqrt{p_1}} |h_{22}|$, we have $\frac{|h_{11}|}{|h_{21}|}\le \frac{ \sqrt{p_2 }}{\sqrt{p_1}}\frac{|h_{22}| }{|h_{21}|}=\frac{ \sqrt{p_2}}{\sqrt{p_1}}(\frac{a_{|\mathcal{S}_{M-1}|+1}}{b_{|\mathcal{S}_{M-1}|+1}}+\frac{|h_{22}| }{b_{|\mathcal{S}_{M-1}|+1}|h_{21}|}) \le \frac{ \sqrt{p_2}}{\sqrt{p_1}}( \frac{a_{k+1}}{b_{k+1}}+\frac{|h_{22}| }{b_{k+1}|h_{21}|})$ for $k=1,\ldots, |\mathcal{S}_{M-1}|$.
	According to Lemma~\ref{thm:subinterval}, we have $g\big(\frac{b_{k}}{a_{k}}, \frac{b_{k+1}}{a_{k+1}}\big)= \frac{b_{k+1} |h_{11}||h_{22}|}{a_{k+1} |h_{21}|+|h_{22}|}  \sqrt{\frac{3 p_1}{M^2-1}}$, for $k=1,\ldots, |\mathcal{S}_{M-1}|$. We also have
	\begin{multline*}
		\max~\Big\{\frac{b_{2} |h_{11}||h_{22}|}{a_{2} |h_{21}|+|h_{22}|}  \sqrt{\frac{3 p_1}{M^2-1}} , \ldots,\\ \frac{b_{|\mathcal{S}_{M-1}|+1} |h_{11}||h_{22}|}{a_{|\mathcal{S}_{M-1}|+1} |h_{21}|+|h_{22}|}  \sqrt{\frac{3 p_1}{M^2-1}}\Big\}
		=|h_{11}| \sqrt{\frac{3 p_1}{M^2-1}},
	\end{multline*}
	with $(w_1^*, w_2^*)=(\sqrt{\frac{3 p_1}{M^2-1}},  \frac{|h_{11}|}{|h_{22}|}\sqrt{\frac{3 p_1}{M^2-1}})$.
	\item If $\frac{|h_{11}|}{|h_{21}| } \in \frac{ \sqrt{p_2}}{\sqrt{p_1}}\big(\frac{a_{\ell_4+1}}{b_{\ell_4+1}}+\frac{|h_{22}| }{b_{\ell_4+1} |h_{21}|}, \frac{a_{\ell_4}}{b_{\ell_4}}+\frac{|h_{22}| }{b_{\ell_4} |h_{21}|}\big)$ for $\ell_4=1,\ldots, |\mathcal{S}_{M-1}|$.
\end{enumerate}
Then, according to Lemma~\ref{thm:subinterval}, we have $g\big(\frac{b_{k}}{a_{k}}, \frac{b_{k+1}}{a_{k+1}}\big)=\frac{b_{k+1} |h_{11}||h_{22}|}{a_{k+1} |h_{21}|+|h_{22}|}  \sqrt{\frac{3 p_1}{M^2-1}}$, for $k=1,\ldots, \ell_4-1$ and  $g\big(\frac{b_{k}}{a_{k}}, \frac{b_{k+1}}{a_{k+1}}\big)=|h_{22}|\sqrt{\frac{3 p_2}{M^2-1}}$ for $k=\ell_4, \ldots, |\mathcal{S}_{M-1}|$.
\begin{align*}
	g\big(\frac{b_{k}}{a_{k}}, \frac{b_{k+1}}{a_{k+1}}\big)\!=\!\begin{cases}
		\frac{b_{k+1} |h_{11}||h_{22}|}{a_{k+1} |h_{21}|+|h_{22}|}  \sqrt{\frac{3 p_1}{M^2-1}} {\rm~for~} k=1,\ldots, \ell_4-1,\\
		|h_{22}|\sqrt{\frac{3 p_2}{M^2-1}}{\rm~for~} k=\ell_4, \ldots, |\mathcal{S}_{M-1}|.
	\end{cases}
\end{align*}
Then, we have,
\begin{align*}
	&\max~\Big\{\frac{b_{2} |h_{11}||h_{22}|}{a_{2} |h_{21}|+|h_{22}|}  \sqrt{\frac{3 p_1}{M^2-1}} , \ldots, \\
	&\qquad \frac{b_{\ell_4} |h_{11}||h_{22}|}{a_{\ell_4} |h_{21}|+|h_{22}|}  \sqrt{\frac{3 p_1}{M^2-1}},  |h_{22}|\sqrt{\frac{3 p_2}{M^2-1}}\Big\}\\
	=&\max~\Big\{\frac{b_{\ell_4} |h_{11}||h_{22}|}{a_{\ell_4} |h_{21}|+|h_{22}|}  \sqrt{\frac{3 p_1}{M^2-1}}, |h_{22}|\sqrt{\frac{3 p_2}{M^2-1}}\Big\}.
\end{align*}
As a result, if $\frac{|h_{11}|}{|h_{21}|}\ge \frac{\sqrt{p_2}}{\sqrt{p_1}}\big(\frac{a_{{\ell_4}}}{b_{\ell_4}}+\frac{|h_{22}| }{b_{\ell_4}|h_{21}|}\big)$, then $\frac{b_{\ell_4} |h_{11}||h_{22}|}{a_{\ell_4} |h_{21}|+|h_{22}|} \sqrt{\frac{3 p_1}{M^2-1}}\ge |h_{22}|\sqrt{\frac{3 p_2}{M^2-1}}$
and $(w_1^*,w_2^*)=\big(\sqrt{\frac{3 p_1}{M^2-1}}, \frac{b_{\ell_4}|h_{11}|}{a_{\ell_4}|h_{21}|+|h_{22}|}\sqrt{\frac{3 p_1}{M^2-1}}\big)$. Else, we can attain $(w_1^*, w_2^*)=\big(\frac{a_{\ell_4+1}|h_{21}|+|h_{22}|}{b_{\ell_4+1} |h_{11}| }  \sqrt{\frac{3 p_2}{M^2-1}}, \sqrt{\frac{3 p_2}{M^2-1}}\big)$. This completes the proof.\hfill$\Box$
\end{appendices}

\small
\bibliographystyle{ieeetr}
\bibliography{tzzt}

\end{document}